\documentclass[fleqn,usenatbib]{mnras} %
\pdfoutput=1

\usepackage{amsmath,amssymb,graphicx}
\usepackage{bm,color}
\usepackage[normalem]{ulem}
\usepackage{mathtools}
\usepackage{url}
\usepackage{footnote}
\makesavenoteenv{table}

\bibpunct{(}{)}{;}{a}{}{,}

\onecolumn

\usepackage{psfrag, epsfig}

\newcommand{\ellbar}{\hbox{\it \l}\,}
\newcommand{\pref}{{\cal A}}
\newcommand{\edth}{\,\eth\,}
\renewcommand{\vec}{\bm}
\renewcommand{\d}[0]{{\rm d}}

\definecolor{darkred}{rgb}{0.8,0.1,0.1}

\newcommand{\Label}[1]{\quad (\mbox{#1})}
\newcommand{\LabelTxt}[1]{(\mbox{#1})}

\definecolor{purple}{RGB}{76, 0,153}

\newcommand{\ontop}[2]{
  \renewcommand{\arraystretch}{0.2}
  \begin{array}{c}
  #1 \\ #2
  \end{array}
  \renewcommand{\arraystretch}{1.0}
}

\newcommand{\gsim}{\ontop{>}{\sim}}

\title[Precision cosmic-shear projection]{Precision calculations of the cosmic shear power spectrum projection}

\author[M.~Kilbinger et~al.]
{Martin Kilbinger$^{1,2}$\thanks{E-mail: martin.kilbinger@cea.fr},
Catherine Heymans$^3$,
Marika Asgari$^3$, 
Shahab Joudaki$^{4,5}$, 
\newauthor
Peter Schneider$^6$,
Patrick Simon$^6$,
Ludovic Van Waerbeke$^7$,
Joachim Harnois-D\'eraps$^3$,
\newauthor
Hendrik Hildebrandt$^6$,
Fabian K\"ohlinger$^8$,
Konrad Kuijken$^9$, and
Massimo Viola$^9$
\\
$^1$CEA/Irfu/SAp Saclay, Laboratoire AIM, 91191 Gif-sur-Yvette, France\\
$^2$Institut d'Astrophysique de Paris, UMR7095 CNRS, Universit\'e Pierre \& Marie Curie, 98 bis boulevard Arago, 75014 Paris, France \\
$^3$Institute for Astronomy, University of Edinburgh, Royal Observatory, Blackford Hill, Edinburgh EH9 3HJ, UK\\
$^4$Centre for Astrophysics \& Supercomputing, Swinburne University of Technology, PO Box 218, Hawthorn, VIC 3122, Australia\\
$^5$ARC Centre of Excellence for All-sky Astrophysics (CAASTRO)\\
$^6$Argelander-Institut f\"ur Astronomie, Auf dem H\"ugel 71, 53121 Bonn, Germany\\
$^7$Department of Physics and Astronomy, University of British Columbia, 6224 Agricultural Road, Vancouver, BC V6T 1Z1, Canada\\
$^8$Kavli Institute for the Physics and Mathematics of the Universe (WPI), The University of Tokyo Institutes for Advanced Study, \\
\phantom{$^7$}The University of Tokyo, Kashiwa, Chiba 277-8583, Japan\\
$^9$Leiden Observatory, Leiden University, Niels Bohrweg 2, 2333 CA Leiden, the Netherlands
}

\date{\today}

\pagerange{\pageref{firstpage}--\pageref{lastpage}} \pubyear{2017}

\begin{document}
\setlength{\voffset}{-12mm}

\label{firstpage}

\maketitle
\begin{abstract}

We compute the spherical-sky weak-lensing power spectrum of the shear and
convergence. We discuss various approximations, such as flat-sky, and first- and
second-order Limber equations for the projection. We find that the impact of
adopting these approximations is negligible when constraining cosmological
parameters from current weak lensing surveys.
This is demonstrated using data from the Canada-France-Hawaii Telescope Lensing Survey
(CFHTLenS). We find that the reported tension with Planck Cosmic Microwave
Background (CMB) temperature anisotropy results cannot be alleviated.
For future large-scale surveys with unprecedented precision, we show that the
spherical second-order Limber approximation will provide sufficient accuracy.
In this case, the cosmic-shear power spectrum is shown to be in agreement with
the full projection at the sub-percent level for $\ell > 3$, with the
corresponding errors an order of magnitude below cosmic variance for all
$\ell$. When computing the two-point shear correlation function, we show that
the flat-sky fast Hankel transformation results in errors below two percent
compared to the full spherical transformation.
In the spirit of reproducible research, our numerical implementation of all
approximations and the full projection are publicly available within the
package \textsc{nicaea} at \texttt{http://www.cosmostat.org/software/nicaea}.

\end{abstract}

\begin{keywords}
cosmological parameters -- methods: statistical
\end{keywords}

\def\aj{AJ}%
\def\araa{ARA\&A}%
\def\apj{ApJ}%
\def\apjl{ApJ}%
\def\apjs{ApJS}%
\def\ao{Appl.~Opt.}%
\def\apss{Ap\&SS}%
\def\aap{A\&A}%
\def\aapr{A\&A~Rev.}%
\def\aaps{A\&AS}%
\def\azh{AZh}%
\def\baas{BAAS}%
\def\jcap{JCAP}%
\def\jrasc{JRASC}%
\def\memras{MmRAS}%
\def\mnras{MNRAS}%
\def\pra{Phys.~Rev.~A}%
\def\prb{Phys.~Rev.~B}%
\def\prc{Phys.~Rev.~C}%
\def\prd{Phys.~Rev.~D}%
\def\pre{Phys.~Rev.~E}%
\def\prl{Phys.~Rev.~Lett.}%
\def\pasp{PASP}%
\def\pasj{PASJ}%
\def\qjras{QJRAS}%
\def\ropp{Rep. Prog. Phys.}%
\def\skytel{S\&T}%
\def\solphys{Sol.~Phys.}%
\def\sovast{Soviet~Ast.}%
\def\ssr{Space~Sci.~Rev.}%
\def\zap{ZAp}%
\def\nat{Nature}%
\def\iaucirc{IAU~Circ.}%
\def\aplett{Astrophys.~Lett.}%
\def\apspr{Astrophys.~Space~Phys.~Res.}%
\def\bain{Bull.~Astron.~Inst.~Netherlands}%
\def\fcp{Fund.~Cosmic~Phys.}%
\def\gca{Geochim.~Cosmochim.~Acta}%
\def\grl{Geophys.~Res.~Lett.}%
\def\jcp{J.~Chem.~Phys.}%
\def\jgr{J.~Geophys.~Res.}%
\def\jqsrt{J.~Quant.~Spec.~Radiat.~Transf.}%
\def\memsai{Mem.~Soc.~Astron.~Italiana}%
\def\nphysa{Nucl.~Phys.~A}%
\def\physrep{Phys.~Rep.}%
\def\physscr{Phys.~Scr}%
\def\planss{Planet.~Space~Sci.}%
\def\procspie{Proc.~SPIE}%
\let\astap=\aap
\let\apjlett=\apjl
\let\apjsupp=\apjs
\let\applopt=\ao

\section{Introduction}
\label{sec:intro}

The measurement of weak gravitational lensing by large-scale structures
provides a powerful cosmological probe of dark matter, dark energy, and
modifications to gravity.  As such it is the primary science goal of several
current (KiDS, DES, HSC\footnote{KiDS: \href={http://kids.strw.leidenuniv.nl},
DES: \href={http://www.darkenergysurvey.org}, HSC:
\href={http://hsc.mtk.nao.ac.jp/ssp}}) and future (Euclid, LSST,
WFIRST\footnote{Euclid: \href={http://sci.esa.int/euclid}, LSST:
\href={http://www.lsst.org}, WFIRST: \href={http://wfirst.gsfc.nasa.gov}})
surveys.
Interest in the results from these surveys is high as statistically significant
deviations have been found between the cosmological parameter constraints from
the CMB Planck experiment \citep{2015arXiv150201589P} in comparison to weak
lensing constraints from both the Kilo-Degree Survey \citep[KiDS;][]{KiDS-450}
and the Canada-France-Hawaii Telescope Lensing Survey
\citep[CFHTLenS;][]{joudaki/etal:2016}.  If the source of this tension is not a
result of so-far unconsidered sources of systematic errors in one or all
experiments, extensions to the standard flat $\Lambda$CDM cosmological models
need to be considered. \citet{joudaki/etal:2017} have shown, for example, that
the tension can be resolved with an evolving dark energy model.

In the era of the upcoming large-scale surveys that will provide measurements
of cosmic shear with unprecedented precision, one needs to revisit the
theoretical predictions of the observables to ensure that the accuracy of the
models meet the accuracy of the observations. In this paper, we examine the
widely used Limber approximation for the projected weak-lensing power spectrum.
We consider spherical coordinates and the flat-sky approximation, and compute
the full projection of the lensing power spectrum. The first-order
extended Limber approximation provides sub-percent accuracy for $\ell > 10$ and
is sufficient for present surveys. The associated errors are sub-dominant even
for future large surveys.

We further show that the second-order Limber approximation is an
accurate representation of the full projection, with better than percent level
precision for scales $\ell > 3$. Since this approximation involves only 1D
integrals over the matter power spectrum, it is very fast to calculate
numerically and can readily be employed in Monte-Carlo sampling methods to
obtain precision constraints on cosmological parameters.
We also compute the shear correlation function using a spherical 
transformation, and compare this to the flat-sky approximated, commonly used
fast Hankel transformation.

This paper is organised as follows.  In section~\ref{sec:wl} we provide a
pedagogical introduction to weak gravitational lensing theory, projections and
power spectra for the flat-sky and spherical cases, following the seminal work
by \citet{2000PhRvD..62d3007H} \cite[see also][]{2005PhRvD..72b3516C}. In
section~\ref{sec:L2} we derive weak-lensing observables using a second-order
Limber approximation first introduced by \citet{2008PhRvD..78l3506L}. We
compare the shear power spectrum and the commonly-used two-point shear
correlation function for the full solution to a range of different
approximations in section~\ref{sec:results}, providing cosmological constraints
for each case using CFHTLenS data from \citet{CFHTLenS-2pt-notomo}. This paper
draws from several sources of literature which have previously discussed the
full projection, or reviewed the accuracy of the Limber approximation for weak
lensing studies, namely \citet{2008PhRvD..78d3002S,2012PhRvD..86b3001B,
2012MNRAS.422.2854G, 2016arXiv161104954K}, see also
\cite{2017arXiv170401054L} for a more recent work. We present a discussion and
comparison of our results to these papers in Appendix~\ref{app:B}.

\section{Weak-lensing projections and power spectra}
\label{sec:wl}

In this section we review the basic weak-lensing projection expressions, and
compute lensing power spectra for a spherical case, and in the flat-sky
approximation. For completeness we provide a derivation of the weak lensing
power spectra in Appendix~\ref{sec:derivations_C}.

\subsection{The lensing potential}
\label{sec:psi}

The lensing potential $\psi$ at a position on the sky $(\theta, \varphi)$ in
the Born approximation is defined as the projected 3D metric potential $\Phi$
along the line of sight of a flat Universe \citep{1998ApJ...498...26K,BS01},
%
%
\begin{equation}
  \psi(\theta, \varphi) = \frac 2 {{\rm c}^2} \int\limits_0^\infty \frac{{\rm d}\chi}{\chi}
    \Phi(\chi, \chi \theta, \chi \varphi; \chi) \, q(\chi),
  \label{eq:psi}
\end{equation}
where the lensing efficiency $q$ is given by
\begin{equation}
  q(\chi) = \int\limits_\chi^{\chi_{\rm h}} {\rm d} \chi^\prime \, n(\chi^\prime)
    \frac{\chi^\prime - \chi}{\chi^\prime},
  \label{eq:lens_efficiency}
\end{equation}
corresponding to a population of lensed galaxies with a normalised source
redshift distribution $n_z(z) {\rm d}z = n(\chi) {\rm d} \chi$, with the limit
being the comoving distance to the horizon $\chi_{\rm h}$\footnote{In
(\ref{eq:psi}) we have replaced this limit without loss of generality with
$\infty$, since $q(\chi > \chi_{\rm h}) = 0$.}. Here, ${\rm c}$ is the speed of
light, and the projection is carried out over comoving distances $\chi$.
The last argument of the potential $\Phi$ is not to
be understood as coordinate, but rather as a substitute of cosmic time,
$t(\chi)$, to express the time at which the potential is evaluated. This is
true in the following for all fields and functions thereof that dynamically
change with cosmic epoch.

The form of the lensing efficiency $q$ in equation (\ref{eq:lens_efficiency})
assumes a homogeneous galaxy distribution without clustering, so that the
redshift distribution in this approximation does not depend on the direction on
the sky. Accounting for this position-dependence leads to corrections of
weak-lensing quantities due to clustering of source galaxies with other sources
\citep{2002A&A...389..729S}, and with galaxies associated to lens structures
\citep{1998A&A...338..375B,H02}.

The 3D potential is related to the density contrast $\delta$ via the Poisson
equation. Assuming General Relativity, this relation is written in Fourier space as
\begin{align}
  \hat \Phi(\vec k; \chi) = & - \frac 3 2 \Omega_{\rm m} H_0^2 k^{-2} a^{-1}(\chi) \hat \delta(\vec k; \chi),
      \label{eq:poisson}
\end{align}
where $\Omega_{\rm m}$ is the matter density parameter, $H_0$ the Hubble
constant, $\vec k$ a 3D Fourier wave vector with modulus $k$ being the comoving
wave number, and $a$ the scale factor with $a=1$ today. The Fourier transform
of the potential and its inverse are defined as
\begin{align}
  \hat \Phi(\vec k; \chi) = &  \int {\rm d}^3 r \, \Phi(\vec r; \chi) {\rm e}^{{\rm i} \vec k \cdot \vec r} ;
  \label{eq:hatPhi}
  \\
  \Phi(\vec r; \chi) = &  \int \frac{{\rm d}^3 k}{(2\pi)^3}
      \hat \Phi(\vec k; \chi) {\rm e}^{-{\rm i} \vec r \cdot \vec k},
  \label{eq:hatPhi_inv}
\end{align}
where the integration range for both integrals is $\mathbb{R}^3$.

\subsection{Lensing power spectra in the spherical case}

\subsubsection{Lensing potential power spectrum}

Following \cite{2000PhRvD..62d3007H} we decompose the lensing
potential (equation~\ref{eq:psi}) into spherical harmonics, in analogy
to the cosmic microwave background (CMB)
temperature, both of which are scalar functions on the sphere. This
decomposition and its inverse are
%
%
\begin{align}
  \psi(\theta, \varphi) = \sum_{\ell=0}^\infty \sum_{m=-\ell}^\ell \psi_{\ell m} {\rm Y}_{\ell m}(\theta, \varphi);
    \label{eq:psi_harm_exp}
    \\
  \psi_{\ell m} = \int_{\mathbb{S}^2} {\rm d} \Omega \, \psi(\theta, \varphi) {\rm Y}^\ast_{\ell m}(\theta, \varphi).
  \label{eq:psi_harm_exp_inv}
\end{align}
Complex conjugation is denoted with the superscript $^\ast$.
To specify tomographic redshift bins $i=1\ldots N_z$, we introduce a family
of lensing efficiency functions $q_i$ defined by a corresponding family of
redshift distributions $n_i$ via equation~(\ref{eq:lens_efficiency}). The resulting
lensing potential is denoted by $\psi_{\ell m, i}$. The tomographic power
spectrum of the lensing potential between two redshift bins $i$ and $j$,
$C_{ij}^\psi$ \citep{pee80} is then defined as
\begin{equation}
  \left\langle \psi^{}_{\ell m, i} \, \psi^\ast_{\ell^\prime m^\prime, j} \right\rangle
    = \delta_{\ell \ell^\prime} \delta_{m m^\prime} C^\psi_{ij}(\ell) .
  \label{eq:C_ell_psi}
\end{equation}
Note that the argument $\ell$ is a discrete integer variable, and is often written as index, $C_\ell$.
Using the properties of the spherical harmonics (see
App.~\ref{sec:derivations_C} for details) the power spectrum can be written as
\begin{align}
  C_{ij}^\psi(\ell) = & \frac 8 {c^4 \pi} 
  \int_0^\infty \frac{{\rm d}\chi}{\chi} q_i(\chi)
  \int_0^\infty \frac{{\rm d}\chi^\prime}{\chi^\prime} q_j(\chi^\prime)
  \int {\rm d} k k^2 \, {\rm j}_\ell(k \chi) {\rm j}_\ell(k \chi^\prime) P_\Phi(k; \chi, \chi^\prime)
  \label{eq:C_ell_phi_Pphi} \\
  = & \frac 8 \pi \pref^2
  \int_0^\infty \frac{{\rm d}\chi}{\chi} \frac{q_i(\chi)}{a(\chi)}
  \int_0^\infty \frac{{\rm d}\chi^\prime}{\chi^\prime} \frac{q_j(\chi^\prime)}{a(\chi^\prime)}
  \int \frac{{\rm d} k}{k^2} \, {\rm j}_\ell(k \chi) {\rm j}_\ell(k \chi^\prime) P_{\rm m}(k; \chi, \chi^\prime) ;
  \label{eq:C_ell_phi_Pm}
\end{align}
where ${\rm j}_\ell$ is the spherical Bessel function of order $\ell$. For
convenience we introduce the normalisation constant $\pref$ as
\begin{equation}
  \pref = \frac 3 2 \Omega_{\rm m} \left(\frac{H_0} c\right)^2.
  \label{eq:pref}
\end{equation}
The first line expresses $C_{ij}^\psi$ in terms of the 3D potential power spectrum $P_\Phi$, defined as
\begin{align}
  \left\langle \hat \Phi(\vec k; \chi) \hat \Phi^\ast(\vec k^\prime; \chi^\prime) \right\rangle
    = & (2\pi)^3 \delta_{\rm D}(\vec k - \vec k^\prime) P_\Phi(k; \chi, \chi^\prime).
  \label{eq:p_phi}
\end{align}
The second line uses the 3D matter density power spectrum $P_{\rm m}$, which is defined analogously as
\begin{align}
  \left\langle \hat \delta(\vec k; \chi) \hat \delta^\ast(\vec k^\prime; \chi^\prime) \right\rangle
    = & (2\pi)^3 \delta_{\rm D}(\vec k - \vec k^\prime) P_{\rm m}(k; \chi, \chi^\prime),
  \label{eq:p_m}
\end{align}
and is related to $P_\Phi$ via the absolute square of the Poisson equation
(\ref{eq:poisson}).

This type of cross-power spectrum between different cosmological epochs
$\chi$ and $\chi^\prime$ was introduced in \cite{2005PhRvD..72b3516C}. In
Sects.~\ref{sec:geom_mean} and~\ref{sec:factor_D} this unequal-time
cross-spectrum \citep{2016arXiv161200770K} will be further evaluated and
simplified in the context of the Limber approximation. The oscillating Bessel
functions in equation~(\ref{eq:C_ell_phi_Pm}) ensure that only relatively close epochs
contribute to the lensing potential correlation. This make sense since observed
light rays from two galaxies at different positions on the sky that necessarily
converge at the observer today, pick up the density fluctuations at similar
times while propagating through the large-scale structure. A similar argument
has been made in \cite{BS01}: since the matter power spectrum scales with $k$
for $k \rightarrow 0$, there is decreasing power towards larger and larger
scales. In particular, the correlation of cosmic fields decreases strongly
above a coherence scale $|\chi - \chi^\prime| \gsim L_{\rm coh}$ which is significantly
smaller than the horizon scale $\chi_{\rm h}$.

In the following section we will discuss the relations between shear and
convergence to the lensing potential on the sphere, and derive the power
spectrum of the former two fields.

\subsubsection{Shear power spectrum}

The shear $\gamma = \gamma_1 + {\rm i} \gamma_2$ is related to the potential at
linear order by the trace-free part of the Jacobi matrix. The involved
differential operator on the sphere is called \emph{edth} derivative, $\edth$,
see \cite{2005PhRvD..72b3516C} for an in-depth mathematical discussion of this
concept. The edth operator $\edth$ ($\edth^\ast$) raises (lowers) the spin $s$
of an object. Twice applying this operator to the scalar (spin-0) potential
creates the spin-2 shear:
\begin{align}
  \gamma(\theta, \varphi) = & \frac 1 2 \edth \edth \psi(\theta, \varphi);
    \nonumber \\
  \gamma^\ast(\theta, \varphi) = & \frac 1 2 \edth^\ast \edth^\ast \psi(\theta, \varphi).
  \label{gamma_psi_spher}
\end{align}
To write the shear on the sphere in terms of the lensing potential $\psi$, we
insert the harmonics expansion of the potential (\ref{eq:psi_harm_exp}). This
requires the calculation of second derivatives of the spherical harmonic functions.
This operation defines a new object, the \emph{spin-weighted spherical
harmonic} $_s{\rm Y}_{\ell m}$. The shear can be written on the sphere in terms
of these functions as a spherical harmonics multipole expansion with
coefficients $_{\pm 2} \gamma_{\ell m}$. This expansion together with its inverse
is
%
%
\begin{align}
  (\gamma_1 \pm {\rm i} \gamma_2)(\theta, \varphi) = & \sum_{\ell m} \,\, _{\pm 2}\gamma_{\ell m} \; _{\pm 2}\!{\rm Y}_{\ell m}(\theta, \varphi);
  \label{eq:gamma_harm_exp}
    \\
  \, _2 \gamma_{\ell m} = & \int_{\mathbb{S}^2} {\rm d} \Omega \, \gamma(\theta, \varphi) \;  _2\!{\rm Y}^\ast_{\ell m}(\theta, \varphi);
    \nonumber \\
  \, _{-2} \gamma_{\ell m} = & \int_{\mathbb{S}^2} {\rm d} \Omega \, \gamma^\ast(\theta, \varphi) \,  _{-2}\!{\rm Y}^\ast_{\ell m}(\theta, \varphi).
  \label{eq:gamma_harm_exp_inv}
\end{align}
The spin-weighted spherical harmonics $_s\!{\rm Y}_{\ell m}$ that are the basis function
in the expansion of the shear (equation~\ref{eq:gamma_harm_exp}) can be calculated via the relations
%
%
\begin{align}
  \ellbar(\ell, s) \; _s\!{\rm Y}_{\ell m}(\theta, \varphi) = & \edth^s {\rm Y}_{\ell m}(\theta, \varphi);
    \nonumber \\
  \ellbar(\ell, s) \; _{-s}\!{\rm Y}_{\ell m}(\theta, \varphi) = & (-1)^s \left(\edth^\ast\right)^s {\rm Y}_{\ell m}(\theta, \varphi),
  \label{eq:sYlm_def} 
\end{align}
for $0 \le s \le \ell$, with the spin pre-factor \citep{2012PhRvD..86b3001B}
\begin{equation}
  \ellbar(\ell, s) = \sqrt{\frac{(\ell + s)!}{(\ell - s)!}}.
  \label{eq:ellbar}
\end{equation} 
Inserting the lensing potential expansion (equation~\ref{eq:psi_harm_exp}) into the
expression for the shear (equation~\ref{gamma_psi_spher}), and using equation~(\ref{eq:sYlm_def})
to compute the derivatives, we find for the shear expansion coefficients
\citep{2000PhRvD..62d3007H,2001astro.ph.11605T}
%
%
\begin{equation}
  _{\pm 2} \gamma_{\ell m} = \frac 1 2 \ellbar(\ell, 2) \psi_{\ell m}.
  \label{eq:gamma_ellm_phi_ellm}
\end{equation}
The two coefficients $_{+2} \gamma_{\ell m}$ and $_{-2} \gamma_{\ell m}$ are
identical since the potential $\psi$ is a real function.

The tomographic shear power spectrum, in analogy to equation~(\ref{eq:C_ell_phi_Pm}), is defined by
\begin{equation}
  \left\langle _2\gamma^{}_{\ell m, i} \; {}_2\gamma^\ast_{\ell^\prime m^\prime, j} \right\rangle
    = \delta_{\ell \ell^\prime} \delta_{m m^\prime} C^\gamma_{ij}(\ell).
  \label{eq:C_ell_gamma}
\end{equation}
%
This is given by
\begin{align}
  C^\gamma_{ij}(\ell) = \frac 1 4 \ellbar^2(\ell, 2) \, C^\psi_{ij}(\ell)
                 = & \frac 2 \pi \, \ellbar^2(\ell, 2) \, \pref^2
                 \int_{0}^\infty \frac{\rm d \chi}{\chi} \frac{q_i(\chi)}{a(\chi)}
                \int_{0}^\infty \frac{\rm d \chi^\prime}{\chi^\prime}
                \frac{q_j(\chi^\prime)}{a(\chi^\prime)}
                \int_0^\infty \frac{{\rm d} k}{k^2} \, P_{\rm m}(k, \chi, \chi^\prime) \,
                {\rm j}_\ell(k \chi) \, {\rm j}_\ell(k \chi^\prime), \Label{FullSph}
  \label{eq:C_ell_full}
\end{align}
where we use the label `FullSph', see Table~\ref{tab:cases} for a list of cases discussed in this work.
The spherical spin pre-factor for the full projection shear power
spectrum is $\ellbar^2(\ell, 2)$, which will be modified under
flat-sky and Limber approximations below.

\subsubsection{Convergence power spectrum}

The convergence is related to the lensing potential on the sphere via the
product of spin-raising and spin-lowering edth operators, which are identical
to the spherical Laplacian differential operator.

\begin{equation}
  \kappa(\theta, \varphi) = \frac 1 2 \edth \edth^\ast \psi(\theta, \varphi) = \frac 1 2 \nabla^2 \psi(\theta, \varphi).
  \label{eq:kappa_psi_spher}
\end{equation}
The spherical harmonics are eigenfunctions of the Laplacian,
\begin{equation}
  \nabla^2 {\rm Y}_{\ell m}(\theta, \varphi) = - \ell (\ell + 1) {\rm Y}_{\ell m}(\theta, \varphi)
    = - \ellbar^2(\ell, 1) {\rm Y}_{\ell m}(\theta, \varphi).
  \label{eq:nabla_Ylm}
\end{equation}
The convergence power spectrum is then similar to the shear power spectrum
(equation~\ref{eq:C_ell_gamma}) with a different spherical pre-factor,
\citep{2000PhRvD..62d3007H,jk12}
\begin{equation}
  C^\kappa_{ij}(\ell) = \frac 1 4 \ellbar^4(\ell, 1) \, C^\psi_{ij}(\ell)
    = \frac{\ell (\ell+1)}{(\ell-1)(\ell+2)} C^\gamma_{ij}(\ell) .
  \label{eq:C_ell_kappa_full}
\end{equation}
The convergence power spectrum is thus larger than the shear power spectrum, by
10\% for $\ell=4$, 1\% for $\ell = 14$, and less than 0.1\% for $\ell>45$.

\subsection{Flat-sky approximation}

The majority of cosmic shear analyses have used the predicted lensing power
spectrum approximated on a flat sky, neglecting the sky curvature. This is a
valid approach for past and current survey areas with an extent less than 10
degrees. To account for the sky curvature of the observed data, the shear
correlation functions from observed galaxy ellipticities are now routinely
computed using spherical coordinates, since projecting to a Cartesian plane has
been shown to cause significant biases of the two-point funtion on large scales
\citep{FSHK08}, and lead to spurious B-modes \citep{asgari/etal:2017}. Here
we examine the effect of sky curvature on the theoretical models of the power
spectrum, and the effect on cosmological parameter inference (see
Sect.~\ref{sec:cfhtlens}).

For a flat-sky, the spherical harmonic expansions are approximated by Fourier
transforms. The flat-sky equivalent of equations~(\ref{eq:psi_harm_exp}) and
(\ref{eq:psi_harm_exp_inv}) are
\begin{align}
  \psi(\vec \vartheta) = & \int \frac{{\rm d}^2 \ell}{(2\pi)^2} \, {\rm e}^{-{\rm i} \vec \ell \cdot \vec \vartheta} \hat \psi(\vec \ell);
  \label{eq:psi_fourier}
  \\
  \hat \psi(\vec \ell) = & \int {\rm d}^2 \vartheta \, {\rm e}^{{\rm i} \vec \ell \cdot \vec \vartheta} \psi(\vec \vartheta),
  \label{eq:psi_fourier_inv}
\end{align}
where $\vec \vartheta = (\theta, \varphi)$ is the vector describing a 2D angle on the sky.
Instead of a harmonics coefficients $\psi_{\ell m}$, the Fourier representation of the potential
$\hat \psi$ now depends on the vector $\vec \ell \in \mathbb{R}^2$.

The flat-sky power spectrum, i.e.\ the flat-sky analogue of equation~(\ref{eq:C_ell_psi}), is defined by
\begin{equation}
  \left\langle \hat \psi_i^{}(\vec \ell) \hat \psi_j^\ast(\vec \ell^\prime) \right\rangle
    = (2\pi)^2 \delta_{\rm D}(\vec \ell - \vec \ell^\prime) P^\psi_{ij}(\ell).
  \label{eq:P_ell_psi}
\end{equation}
\cite{2000PhRvD..62d3007H} has shown that for small angles the harmonics
expansion (equation~\ref{eq:psi_harm_exp}) can be approximated by the Fourier
representation in equation~(\ref{eq:psi_fourier}). They also demonstrated that the power
spectra are approximately equal, $P^\psi \approx C^\psi$.

For a spin-2 field, \cite{2000PhRvD..62d3007H} approximates the edth operator
by Cartesian derivatives, and approximates equation~(\ref{eq:sYlm_def}) as
\begin{equation}
  \ell^2 \, _{\pm2}\!{\rm Y}_{\ell m}(\theta, \varphi) \approx  {\rm e}^{{\mp}2 {\rm i} \phi_\ell}
    (\partial_1 \pm {\rm i} \partial_2)^2 \, {\rm Y}_{\ell m}(\theta, \varphi).
  \label{eq:sYlm_def_flat}
\end{equation}
The spin pre-factor $\ellbar(\ell, 2) = \sqrt{(\ell-1) \ell (\ell+1) (\ell+2)}$
is replaced by $\ell^2$ in flat co-ordinates, an approximation that holds for
large $\ell$, since sky curvature can be neglected for small angular
scales. We find for the flat-sky shear power spectrum
\begin{align}
  P^\gamma_{ij}(\ell) 
                 = & \frac 2 \pi \, \ell^4 \, \pref^2
                 \int_{0}^\infty \frac{\rm d \chi}{\chi} \frac{q_i(\chi)}{a(\chi)}
                \int_{0}^\infty \frac{\rm d \chi^\prime}{\chi^\prime}
                \frac{q_j(\chi^\prime)}{a(\chi^\prime)}
                \int_0^\infty \frac{{\rm d} k}{k^2} \, P_{\rm m}(k, \chi, \chi^\prime) \,
                {\rm j}_\ell(k \chi) \, {\rm j}_\ell(k \chi^\prime) .
  \label{eq:P_ell_gamma_full_der}
\end{align}
with flat-sky pre-factor $\ell^4$. See App.~\ref{sec:schmidt08} and~\ref{sec:giannantonio12} for discussions of alternative expressions for the
flat-sky power spectrum.

\section{Second-order Limber approximation for weak lensing}
\label{sec:L2}

\subsection{Spherical case}

We follow \cite{2008PhRvD..78l3506L} who derive the second-order Limber
expansion for general projections from 3D to 2D scalar fields in the spherical,
all-sky case. We apply their general derivation to the case of weak lensing,
and contrary to \cite{2008PhRvD..78l3506L} account for a time-dependent power
spectrum using two approaches presented in Sects.~\ref{sec:geom_mean} and~\ref{sec:factor_D}.

First, we use the identity of Bessel functions
\begin{equation}
  {\rm j}_\ell(x) = \sqrt{\frac{\pi}{2x}} {\rm J}_{\ell + 1/2}(x)
  \label{eq:jJ}
\end{equation}
in equation~(\ref{eq:C_ell_full}), where ${\rm J}_\nu$ is the Bessel function of the first kind
and order $\nu$. Next, \cite{2008PhRvD..78l3506L} solve integrals of
the form
\begin{equation}
  \int_0^\infty {\rm d} \chi f(\chi) {\rm J}_\nu(k \chi)
  = \int_0^\infty {\rm d} x k^{-1} f(x/k) {\rm J}_\nu(x)
  \label{eq:int_dchi}
\end{equation}
by performing a Taylor expansion of an arbitrary differentiable function
$f$ around $x = k \chi = \nu = \ell + 1/2$, where the Bessel function has its
approximate maximum.

\subsubsection{Geometric mean cross-correlation power spectrum}
\label{sec:geom_mean}

To separate the $k$- and $\chi, \chi^\prime$-terms in
equation~(\ref{eq:C_ell_full}), we first approximate the matter power
cross-spectrum between two distances by the geometric mean of the two involved
distances \citep{2005PhRvD..72b3516C,2016arXiv161200770K},
\begin{equation}
 P_{\rm m}(k; \chi, \chi^\prime) = \sqrt{ P_{\rm m}(k; \chi) P_{\rm m}(k; \chi^\prime) } \, .
  \label{eq:P_geom_mean}
\end{equation}
This form is justified when considering the linear power spectrum, and
follows when inserting the linearly evolving density contrast $\hat \delta(\vec
k; \chi) = D_+(\chi) \hat \delta_0(\vec k)$ into equation~(\ref{eq:p_m}), where
$\delta_0$ is the present-day linearly extrapolated density contrast, and $D_+$
the linear growth factor with $D_+(0) = 1$. This is a good approximation
also in the non-linear case as shown in \cite{2016arXiv161200770K}.

With this, equation~(\ref{eq:C_ell_full}) is written as
\begin{align}
  C^\gamma_{ij}(\ell) \approx & \, \ellbar^2(\ell, 2) \, \pref^2
                \int_0^\infty \frac{{\rm d} k}{k^3} \,
                \int_{0}^\infty \frac{{\rm d} \chi}{\chi^{3/2}} \sqrt{P_{\rm m}(k; \chi)}
                \frac{q_i(\chi)}{a(\chi)} {\rm J}_{\ell+1/2}(k \chi)
                \int_{0}^\infty \frac{{\rm d} \chi^\prime}{{\chi^\prime}^{3/2}}
                \sqrt{P_{\rm m}(k; \chi^\prime)} \frac{q_j(\chi^\prime)}{a(\chi^\prime)} {\rm J}_{\ell+1/2}(k \chi^\prime) \, .
  \label{eq:C_ell_full_Pk}
\end{align}
Note that this equation has a pre-factor $\ellbar^2(\ell, 2)$ corresponding to a spin-2 field, in contrast
to \cite{2008PhRvD..78l3506L} who show calculations for a scalar field.

Following \cite{2008PhRvD..78l3506L} we expand to third order
\begin{equation}
  \lim_{\varepsilon \rightarrow 0} \int_0^\infty {\rm d} x \, {\rm e}^{-\epsilon (x - \nu)} g(x) {\rm J}_\nu(x)
  \approx g(\nu) - \frac 1 2 \left.\frac{{\rm d}^2 g}{{\rm d} x^2}\right|_{x=\nu}
                 - \frac \nu 6 \left.\frac{{\rm d}^3 g}{{\rm d} x^3}\right|_{x=\nu},
\end{equation}
with $g(x) = k^{-1} f(k, \chi)$, $\chi=x/k$, and its derivatives $g^{(n)}(x) =
k^{-1-n} f^{(n)}(k, \chi)$ for a given $k$, where the derivatives of $f$
are with respect to the second argument $\chi$. In this series expansion, the
replacement of the integral with the evaluation of $g$ and its derivatives at
the maximum of the Bessel function is a good approximation of the integral if
$g$ is varying more slowly than the oscillating Bessel function.
As we will show below, $f$ is a slowly varying function of the comoving
distance. In our case the projection kernel is
\begin{equation}
  f(k, \chi) = \sqrt{P_{\rm m}(k; \chi)} \, a^{-1}(\chi) \chi^{-3/2} q(\chi).
  \label{eq:f_LA08}
\end{equation}
In the tomographic case, the index $i$ is added to $q$ and $f$.
Replacing both distance integrals in equation~(\ref{eq:C_ell_full_Pk}) by their
Taylor-expansions around the maxima $\nu(\ell) = \ell + 1/2$ of the two Bessel
functions, which are $k \chi$ and $k \chi^\prime$, respectively, yields
\begin{align}
  C^\gamma_{ij}(\ell) \approx & \, \ellbar^2(\ell, 2) \, \pref^2
    \int_0^\infty \frac{{\rm d} k}{k^3} \, k^{-2}
    \left[ f_i(k, \chi) - \frac{1}{2 k^2} f_i^{\prime\prime}(k, \chi)
      - \frac{\nu(\ell)}{6 k^3} f_i^{\prime\prime\prime}(k, \chi) + \ldots \right]
    \left[ f_j(k, \chi) - \frac{1}{2 k^2} f_j^{\prime\prime}(k, \chi)
    - \frac{\nu(\ell)}{6 k^3} f_j^{\prime\prime\prime}(k, \chi) + \ldots \right] \, .
  \label{eq:C_ell_limber2_dk}
\end{align}
Changing the integration to $\chi = \nu/k$ and collecting terms according to their $\nu$-dependence:
\begin{align}
  C^\gamma_{ij}(\ell) \approx & \, C^\gamma_{{\rm L1}, ij}(\ell) + C^\gamma_{{\rm L2}, ij}(\ell)
    = 
    \frac{\ellbar^2(\ell, 2)}{\nu^4(\ell)} \, \pref^2
    \int_0^\infty {\rm d} \chi \, \chi^3 \, 
    \Bigg\{
    (f_i f_j)(\nu(\ell)/\chi, \chi)
      \nonumber \\
    &
     - \frac 1 {\nu^2(\ell)}
    \left[ \frac{\chi^2}{2} \left( f^{}_i f^{\prime\prime}_j + f_i^{\prime\prime} f^{}_j \right)(\nu(\ell)/\chi, \chi)
     + \frac{\chi^3}{6} \left( f^{}_i f^{\prime\prime\prime}_j + f^{\prime\prime\prime}_i f^{}_j \right)(\nu(\ell)/\chi, \chi)
    \right]
    + {\cal O}(\nu^{-4})
    \Bigg\}
    \, .
  \label{eq:C_ell_limber12_dr}
\end{align}
The first term corresponds to the well-known first-order Limber approximation
\citep{1953ApJ...117..134L,1992ApJ...388..272K}, which is widely used in
weak gravitational lensing. We retrieve the (spherical) standard expression by
inserting back the projection kernel (equation~\ref{eq:f_LA08}),
\begin{align}
  C^\gamma_{{\rm L1}, ij}(\ell) = & \, \frac{\ellbar^2(\ell, 2)}{\nu^4(\ell)} \, \pref^2 \int {\rm d} \chi \frac{ q_i(\chi) q_j(\chi) }{a^2(\chi)}
  P_{\rm m}\left(\frac{\nu(\ell)}{\chi}; \chi\right) . \Label{ExtL1Sph}
  \label{eq:C_ell_limber1}
\end{align}
In the Limber approximation, modes between structures at different epochs do
not contribute to the single line-of-sight integration.

The second-order Limber term in equation~(\ref{eq:C_ell_limber12_dr}) has an additional
$\nu^{-2}$-dependence, and is therefore strongly suppressed for large $\ell$,
\begin{align}
  C^\gamma_{{\rm L2}, ij}(\ell) = & - \frac 1 {\nu^2(\ell)} \, \frac{\ellbar^2(\ell, 2)}{\nu^4(\ell)} \,
    \frac{\pref^2}{2}
    \int {\rm d} \chi \, \chi^{7/2} a^{-1} P^{1/2}_{\rm m}\left(\frac{\nu(\ell)}\chi; \chi\right)
    \left[ q^{}_i f^{\prime\prime}_j + f^{\prime\prime}_i q^{}_j
      + \frac{\chi}{3} \left( q^{}_i f^{\prime\prime\prime}_j + f^{\prime\prime\prime}_i q^{}_j
      \right)
    \right](\nu(\ell)/\chi, \chi) \, . 
  \label{eq:C_ell_limber2_dr} 
\end{align}
The higher-order derivatives of the filter functions have to be computed
numerically in the general case. These suffer from numerical noise and are
sensitive to the set-up, for example the step size. The tabulation and
interpolation of those derivatives is time-consuming since they depend on two
arguments, $\nu$ and $\chi$. In the following section, we will separate the
$k$- and $\chi$-dependent parts of the power spectrum to make the numerical
derivatives faster and more smooth.

\subsubsection{Approximation for small $\ell$}
\label{sec:factor_D}

To further develop equation~(\ref{eq:P_geom_mean}), we divide out the growth factor of
the power spectrum,
\begin{equation}
 P_{\rm m}(k, \chi) =: D^2_+(\chi) \tilde P_{\rm m}(k, \chi) \approx D^2_+(\chi) \tilde P_{\rm m}(k).
  \label{eq:P_k_chi_sep_s}
\end{equation}
The function $\tilde P_{\rm m}$ is in general not time-independent, except
in the case of a linear matter power spectrum in the absence of massive
neutrinos, and in General Relativity. However, the second-order Limber terms
are expected to be important only for small $\ell$, since for large $\ell$ the first-order
Limber equation (\ref{eq:C_ell_limber1}) is dominant. Eq.~(\ref{eq:P_k_chi_sep_s})
becomes a good approximation for small $\ell$, since that means either small
$k$ where the evolution is linear, or small $\chi$, where the lensing
efficiency is small. The accuracy of a very nearby tomographic bin with low
mean redshift should be further examined, but this is not the case for
CFHTLenS.

With the definition (\ref{eq:P_k_chi_sep_s}), we factor out the function
$\tilde P_{\rm m}(k, \chi)$ from equation~(\ref{eq:f_LA08}), and we define the
separated kernel function $f_{\rm s}$ as
\begin{equation}
  f_{\rm s}(\chi) = D_+(\chi) a^{-1}(\chi) \chi^{-3/2} q(\chi) \, .
  \label{eq:f_LA08_s}
\end{equation}
The Limber equation up to second order of the shear power spectrum can then be approximated as
\begin{align}
  C^\gamma_{ij}(\ell) = & \, C^\gamma_{{\rm L1}, ij}(\ell) 
    \nonumber \\
    & - \frac 1 {\nu^2(\ell)} \, \frac{\ellbar^2(\ell, 2)}{\nu^4(\ell)} \,
    \frac{\pref^2}{2}
    \int {\rm d} \chi \, \chi^{7/2} a^{-1}(\chi) D_+^{-1}(\chi) P_{\rm m}\left(\frac{\nu(\ell)}\chi; \chi\right)
    \left[ q^{}_i f^{\prime\prime}_{{\rm s} j} + f^{\prime\prime}_{{\rm s}i} q^{}_j
      + \frac{\chi}{3} \left(
        q^{}_i f^{\prime\prime\prime}_{{\rm s}j} + f^{\prime\prime\prime}_{{\rm s}i} q^{}_j
      \right)
    \right](\chi) \, . \Label{ExtL2Sph}
  \label{eq:C_ell_limber2_dr_s}
\end{align}
We compute the numerical higher derivatives as follows. The functions $f_{\rm
s}(\chi)$ are fitted as power laws with index $\approx -1.5$, which is
expected if the growth suppression factor $D_+(a)/a$ varies slowly with $\chi$,
and the lensing efficiency $q \approx 1$ for small and medium $\chi$, given the
CFHTLenS redshift range. The fit is carried out between $\chi_{\rm min} = 0.001
\, {\rm Mpc}/h$ and $\chi_{\rm max} = 500 \, \mbox{Mpc}/h$. At larger comoving
distances the kernel decreases faster than a power law, so we exclude this
range from the fit. Even though on those scales the derivatives are larger due
to the steeper decline, the associated errors are very small as these scales
are down-weighed by the kernel function $f_{\rm s}$ itself. At $\chi = 500 \,
\mbox{Mpc}/h$ the filter function is less than $10^{-4}$ of its value 
at $1$\,Mpc$/h$.

\subsection{Flat-sky}
\label{sec:Limber_flat_sky}

The extended flat-sky Limber approximation is readily derived from the
spherical case equations~(\ref{eq:C_ell_limber1},~\ref{eq:C_ell_limber2_dr_s}), by
replacing the pre-factor $\ellbar^2(\ell, 2)$ with $\ell^4$,
\begin{align}
  P^\gamma_{{\rm L1}, ij}(\ell) = & \, p(\ell) \, \pref^2 \int {\rm d} \chi \frac{ q_i(\chi) q_j(\chi) }{a^2(\chi)}
  P_{\rm m}\left(\frac{\nu(\ell)}{\chi}; \chi\right);
  \label{eq:P_ell_limber1}
  \\
    P^\gamma_{{\rm L2}, ij}(\ell) = & - \frac 1 {\nu^2(\ell)} \, p(\ell) \, \frac{\pref^2}{2}
    \int {\rm d} \chi \, \chi^{7/2} a^{-1}(\chi) D_+^{-1}(\chi) P_{\rm m}\left(\frac{\nu(\ell)}\chi; \chi\right)
    \left[ q^{}_i f^{\prime\prime}_{{\rm s}j} + f^{\prime\prime}_{{\rm s}i} q^{}_j  
      + \frac{\chi}{3} \left( q^{}_i f^{\prime\prime\prime}_{{\rm s}j} + f^{\prime\prime\prime}_{{\rm s}i} q^{}_j
      \right)
    \right](\chi)
  \label{eq:P_ell_limber2}
\end{align}
Further approximations can be made for the pre-factor $p(\ell) = \ell^4/\nu^4(\ell)$ and $\nu(\ell)$:
\begin{enumerate}
  \item $p(\ell) = 1$, this corresponds to $\nu(\ell) = \ell$, which is the
    \emph{standard} Limber approximation \LabelTxt{L1Fl} Until recently,
  i.e.~for all pre-2014 CFHTLenS results and DLS
  \citep{2012arXiv1210.2732J} analyses, this was the approximation of choice. Note that
  we do not discuss the second-order Limber approximation with $p(\ell)=1$.
  \item $p(\ell) = \ell^4/(\ell + 1/2)^4$. This corresponds to the
  \emph{extended} Limber approximation \LabelTxt{ExtL1Fl, ExtL2Fl} with
  $\nu(\ell) = \ell + 1/2$; however the following case is typically employed:
  \item $p(\ell) = 1$, but keeping as argument of the power spectrum $\nu(\ell) = \ell + 1/2$. This is
    a \emph{hybrid} between
    standard and extended Limber approximation~\LabelTxt{ExtL1FlHyb, ExtL2FlHyb}, and the first-order case 
    was used in \cite{KiDS-450,joudaki/etal:2016,joudaki/etal:2017,abbott/etal:2016}. As is shown
    below, this is a better approximation to the full projection than case (ii). In equation~(\ref{eq:P_ell_limber2})
    the second-order suppression factor is also left to be $\nu^{-2}(\ell) = (\ell + 1/2)^{-2}$, providing
    a slightly more accurate approximation compared to $\nu^{-2}(\ell) = \ell^{-2}$.
\end{enumerate}

\section{Results}
\label{sec:results}

\subsection{Comparison of the approximations for the lensing power spectrum}
\label{sec:comp}

\begin{figure}

  \begin{center}
    \resizebox{1.0\hsize}{!}{
      \includegraphics{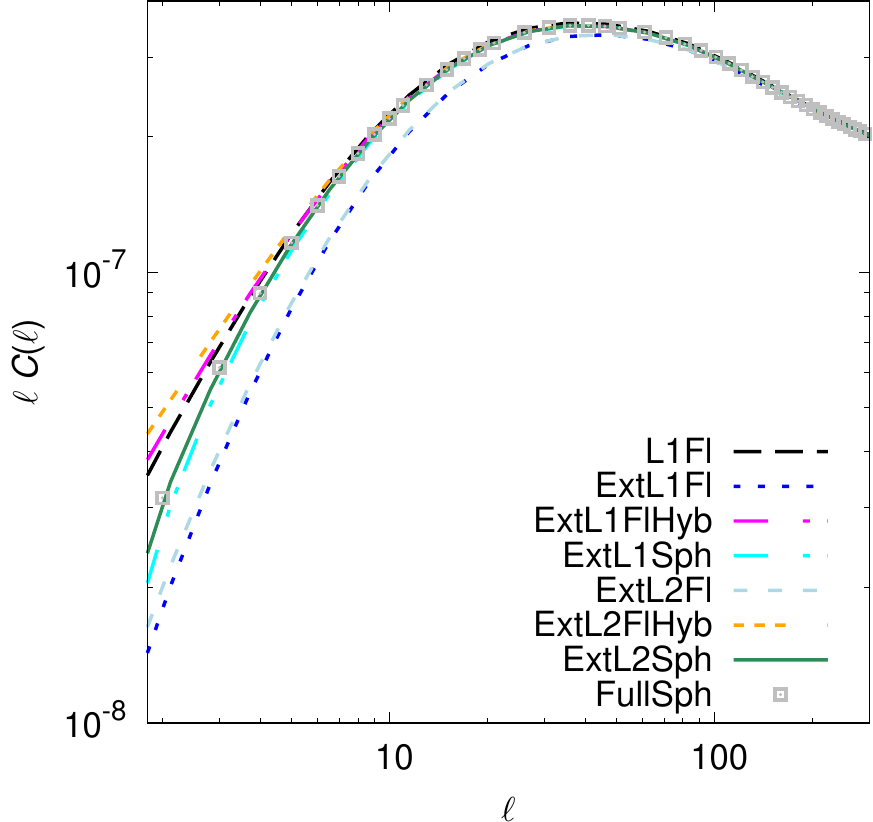}
      \hspace*{2em}
      \includegraphics{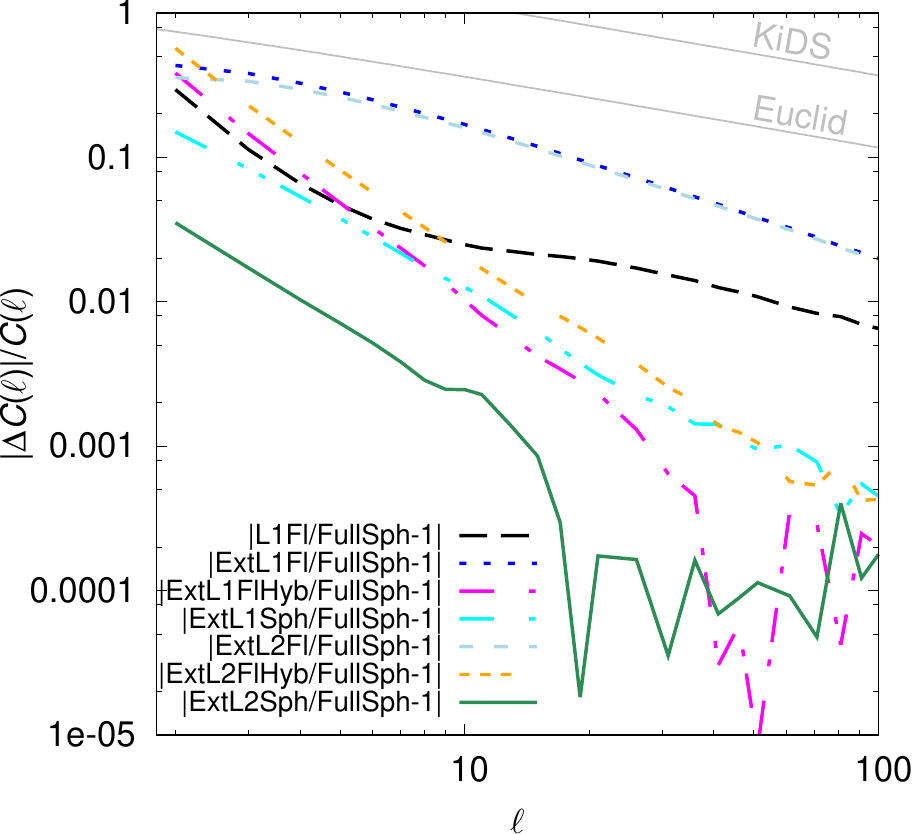}
    }
  \end{center}

  \caption{\label{fig:Cl_cases}%
        The shear power spectrum for different approximations listed in Table~\ref{tab:cases}.
        Limber to first order: standard with flat-sky (L1Fl),
        extended for flat sky (ExtL1Fl), extended hybrid for flat sky (ExtL1FlHyb),
        and extended in the spherical expansion (ExtL1Sph);
        second-order Limber approximations: extended flat sky (ExtL2Fl), extended hybrid flat sky (ExtL2FlHyb),
        and extended spherical expansion (ExtL2Sph); full (exact) spherical projection (FullSph).
        The left panel shows the total shear power spectrum.  The right panel shows the fractional difference
        resulting from each approximation, 
        relative to the full spherical projection of the shear power spectrum. The two light grey curves on the
        top show the cosmic variance for KiDS- and Euclid-like surveys with areas of $1,500$ and $15,000$
        square degrees, respectively.
        }
\end{figure}

In Fig.~\ref{fig:Cl_cases} we present the full spherical projection of the
shear power spectrum in comparison to shear power spectra derived assuming the
range of different approximations listed in Table~\ref{tab:cases}.  The adopted
redshift distribution corresponds to CFHTLenS \citep{CFHTLenS-2pt-notomo} and
we assume their best-fit flat $\Lambda$CDM cosmology with $\Omega_{\rm
m}=0.279$, $\Omega_{\rm b}=0.046$, $\sigma_8=0.79$, $h=0.701$, $n_{\rm
s}=0.96$. For $\ell > 100$ we find that all shear power spectra predictions
agree with the full spherical solution to better than one percent, with the
majority of the approximations tested accurate to better than 0.1 per cent.   

Considering first the flat-sky cases, the standard first-order Limber
approximation, (L1Fl), that was adopted for all pre-2014 CFHTLenS analyses and
DLS analyses, we find it to be accurate to better than 10\% for $\ell>3$,
converging slowly to the true projection with percent level precison at
$\ell>100$. For the extended Limber approximations `hybrid' cases (ExtL1FlHyb
and ExtL2FlHyb), despite decreased accuracy for $\ell < 8$ in comparison to the
standard first-order Limber case, the errors with respect to the true power
spectrum decrease much faster, as $\ell^{-2}$, such that percent-level
precision is reached at $\ell>15$.  The first-order extended Limber
approximation `hybrid' cases (ExtL1FlHyb) was adopted by
\citet{joudaki/etal:2016}, \citet{joudaki/etal:2017}, DES-SV
\citep{abbott/etal:2016} and \cite{KiDS-450}\footnote{We confirm that there is
a typographical error in equation 4 of \cite{KiDS-450} and in equation 2 of
\citet{abbott/etal:2016} (private communication with Joe Zuntz) which should
include the extra term of `+0.5' in $\nu(\ell)$ that was incorporated in both
cosmological analyses.}. 

The outlier in the flat-sky cases is the extended Limber approximation
(ExtL1Fl) which performs relatively poorly, and reaches 10\% precision only at
$\ell > 100$. The same slow convergence can be observed for the corresponding
second-order flat case, ExtL2Fl. To our knowledge this form of the flat-sky
approximation has not been used in any cosmic shear studies to date, and should
not be used in any future studies given these results.  The poor behaviour of
this case, in comparison with the `hybrid' case, (e.g.~ExtL1FlHyb) can be
understood by considering Taylor expansions of the different pre-factors.  The
spherical pre-factor,
\begin{equation}
p(\ell) = \frac{(\ell+2)(\ell+1)\ell(\ell-1)}{(\ell + 0.5)^4} = 1-{5\over 2\ell^2} +{\cal O}(\ell^{-3}) \, ,
\end{equation}
can be compared to the flat-sky extended Limber pre-factor
\begin{equation}
p(\ell) = \ell^4 / (\ell + 0.5)^4 = 1-{2\over \ell}+{5\over 2\ell^2} +{\cal O}(\ell^{-3}) \, . 
\end{equation} 
showing it to be more deviant from the full spherical solution, than the
`hybrid' $p(\ell) = 1$ case. 

Considering now the spherical-sky cases, we find that including the spherical
pre-factor decreases the difference between the extended Limber approximated
cases (ExtL1Sph and ExtL2Sph) and the full spherical solution by a factor of a
few for $\ell < 5$.   We find that using the spherical-sky second-order
extended Limber approximation (ExtL2Sph) yields percent accuracy down to $\ell
> 3$.  The numerical calculation of the second-order extended Limber
approximation is a factor of 15 times faster than the calculation of the
full spherical solution (averaged over the first $18$ $\ell$-modes). We
note that the sub-$0.1\%$-fluctuations seen in the right panel of
Fig.~\ref{fig:Cl_cases} is due to numerical noise arising from numerical
integration errors in the calculation of the full spherical solution when $\ell
> 20$.

We note that in all cases the second-order Limber expansion adds power
to the first-order term. In the flat hybrid case, which over-estimates the full
spherical solution, this results in the second-order expansion being less
precise compared to first-order.

Compared to the statistical power of current and future surveys, all
approximations discussed here are subdominant to the cosmic variance, $\Delta
C(\ell) / C(\ell) = [f_{\rm sky} (2 \ell + 1)]^{-1/2}$
\citep{1992ApJ...388..272K}, where $f_{\rm sky}$ is the fraction of sky observed
by the survey. The uncertainties from the Limber approximation in the case of
ExtL2Sph is an order of magnitude below the cosmic variance of a Euclid-like
survey (sky area $15,000$ square degrees) for all $\ell$.

\renewcommand{\baselinestretch}{1.5}
\begin{table}
\begin{centering}
  \caption{\label{tab:cases}The shear power spectrum approximations studied in this paper. `ID' is the label
    used in the text and figures.
    The sixth (seventh) column indicates the mode $\ell_x$ such that for $\ell \leq \ell_x$, the approximated
    power spectrum is more accurate than $x$, with $x=0.1$ ($0.01$).
    } 
    \begin{tabular}{|p{0.22\textwidth}|l|c|c|c|c|c|p{0.21\textwidth}}
  \hline
  Case & ID & equation & $p(\ell)$ & $\nu(\ell)$ & $\ell_{0.1}$ & $\ell_{0.01}$ & Comment \\ \hline
  $1^{\rm st}$-order standard Limber, flat & L1Fl & (\ref{eq:P_ell_limber1})+(i)
    & $1$ & $\ell$ & $4$ & $60$ & Pre-2014 CFHTLenS and DLS \\ \hline
  $1^{\rm st}$-order extended Limber, flat & ExtL1Fl & (\ref{eq:P_ell_limber1})+(ii)
    & $\frac{\ell^4}{(\ell+1/2)^4}$ & $\ell + \frac 1 2$ & $20$ & $200$ & Converges only with ${\cal O}(\ell^{-1})$ \\ \hline
  $1^{\rm st}$-order extended Limber, hybrid, flat & ExtL1FlHyb & (\ref{eq:P_ell_limber1})+(iii)
    & $1$ & $\ell + \frac 1 2$ & $4$ & $10$ & Post-2014 CFHTLenS, DES-SV and KiDS \\ \hline
  $1^{\rm st}$-order extended Limber, spherical & ExtL1Sph & (\ref{eq:C_ell_limber1})
    & $\frac{\ellbar^2(\ell, 2)}{(\ell+1/2)^4}$ & $\ell+ \frac 1 2$ & $3$ & $12$ & \\ \hline
  $2^{\rm st}$-order extended Limber, flat & ExtL2Fl &  (\ref{eq:P_ell_limber1})+(\ref{eq:P_ell_limber2})+(iii)
    & $\frac{\ell^4}{(\ell+1/2)^4}$ 
    & $\ell+\frac 1 2$ & $19$ & $200$ & Converges only with ${\cal O}(\ell^{-1})$ \\ \hline
  $2^{\rm st}$-order extended Limber, hybrid, flat & ExtL2FlHyb &  (\ref{eq:P_ell_limber1})+(\ref{eq:P_ell_limber2})+(iii)
    & $1$ 
    & $\ell+ \frac 1 2$ & $5$ & $16$ & Best flat-sky approximation \\ \hline
  $2^{\rm st}$-order extended Limber, spherical & ExtL2Sph & (\ref{eq:C_ell_limber1})+(\ref{eq:C_ell_limber2_dr_s})
    & $\frac{\ellbar^2(\ell, 2)}{(\ell+1/2)^4}$ & $\ell+\frac 1 2$ & $2$ & $5$ & Best approximation \\ \hline
  Full spherical & FullSph & (\ref{eq:C_ell_full}) &
      - & - & - & - & Correct projection \\ \hline
  \end{tabular}

\end{centering}
\end{table}
\renewcommand{\baselinestretch}{1}

\subsection{Effects on the shear correlation function}
\label{sec:comp_xi}

The majority of cosmic shear analyses to date have adopted real-space
correlation statistics, since these can be measured directly from an observed
galaxy shape catalogue. The baseline quantity is the two-point correlation
function \citep{1991ApJ...370....1M, 1992ApJ...388..272K, BS01}, given in the flat-sky approximation by
\begin{align}
  \xi_+(\theta) 
  = \left\langle \gamma \gamma^\ast \right\rangle(\theta) = \frac 1 {2\pi} \int {\rm d} \ell \, \ell \, {\rm J}_0(\ell
   \theta)
  P^\gamma(\ell);
  \quad
   %
   \xi_-(\theta)
  = \left\langle \gamma \gamma \right\rangle(\theta) = \frac 1 {2\pi} \int
   {\rm d} \ell \, \ell \, {\rm J}_4(\ell \theta)
  P^\gamma(\ell).
   \label{eqn:xiGG}
\end{align}
The flat-sky shear power spectrum $P^\gamma$ can be related to the underlying
matter power spectrum through equation~(\ref{eq:P_ell_limber1}) when adopting a
first-order extended Limber approximation, or equation~(\ref{eq:P_ell_limber2})
when adopting a second-order extended Limber approximation.

On a sphere, correlation functions formally cannot be related to the power
spectrum by the Hankel transform in equation~(\ref{eqn:xiGG}), and should be
replaced by the spherical transform
\citep{1999IJMPD...8...61N,2004MNRAS.350..914C}
\begin{equation}
 \xi_+(\theta) 
  =  \frac 1{4\pi} \sum\limits_{\ell=2}^\infty (2 \ell + 1) C^\gamma(\ell) {\rm d}^\ell_{2\,2}(\theta); \quad 
 \xi_-(\theta) 
  =  \frac 1{4\pi} \sum\limits_{\ell=2}^\infty (2 \ell + 1) C^\gamma(\ell) {\rm d}^\ell_{2\,-2}(\theta)
  \label{eq:xi_wigner}
\end{equation}
where ${\rm d}^\ell_{m\,n}$ are the reduced Wigner ${\rm D}$-matrices, see App.~\ref{app:C} for details
on their numerical calculation.

The spherical power spectrum is formally not defined for non-integer $\ell$
\citep[see][for alternative spherical-sky formulae for the two-point
correlation function]{2005PhRvD..72b3516C}, as functions defined on the
sphere are necessarily periodic.
As we have shown in section~\ref{sec:comp}, however, the spherical second-order
extended Limber approximation provides a percent-level precision representation
of the full spherical projection for $\ell > 3$.
The spherical pre-factor $\ellbar(\ell, 2)$ (equation~\ref{eq:ellbar}) can be
generalised to non-integer arguments and is positive for $\ell \ge 2$. We can
thus use the spherical power spectrum with
the Hankel transformation in equation~(\ref{eqn:xiGG}) to compute the two-point correlation
functions. This has the advantage being able to employ fast FFT numerical implementations of
the Hankel tranforms \citep{2000MNRAS.312..257H}, when Monte-Carlo sampling.

\begin{figure}

  \begin{center}
    \resizebox{1.0\hsize}{!}{
      \includegraphics{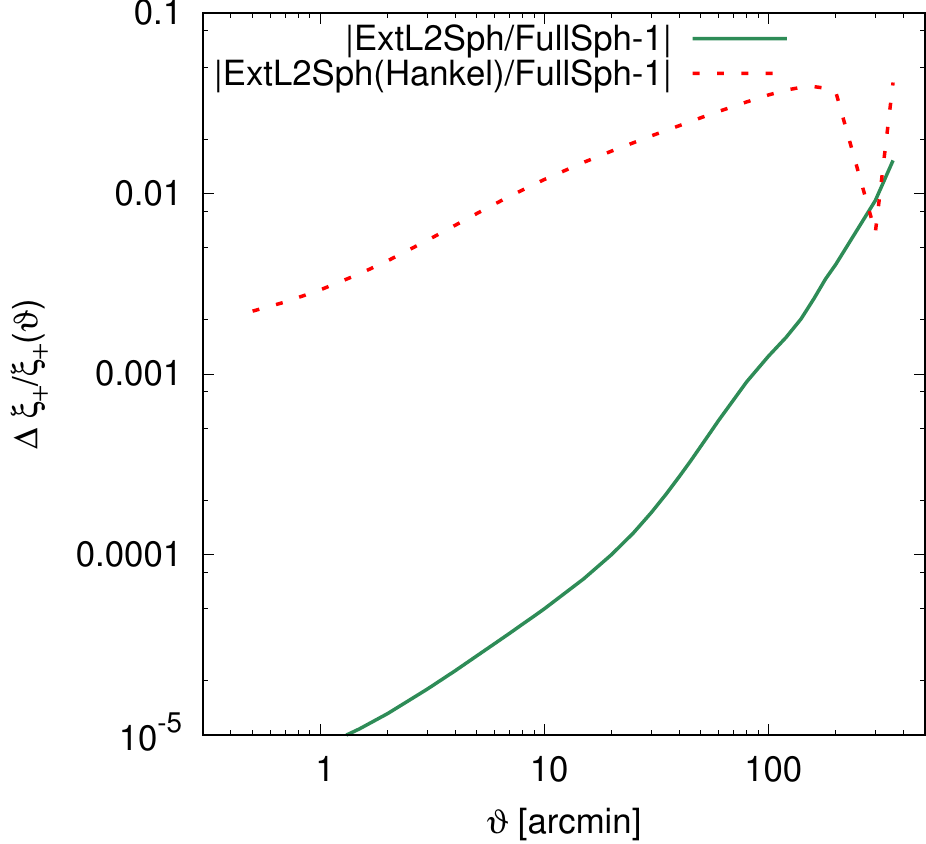}
      \hspace*{2em}
      \includegraphics{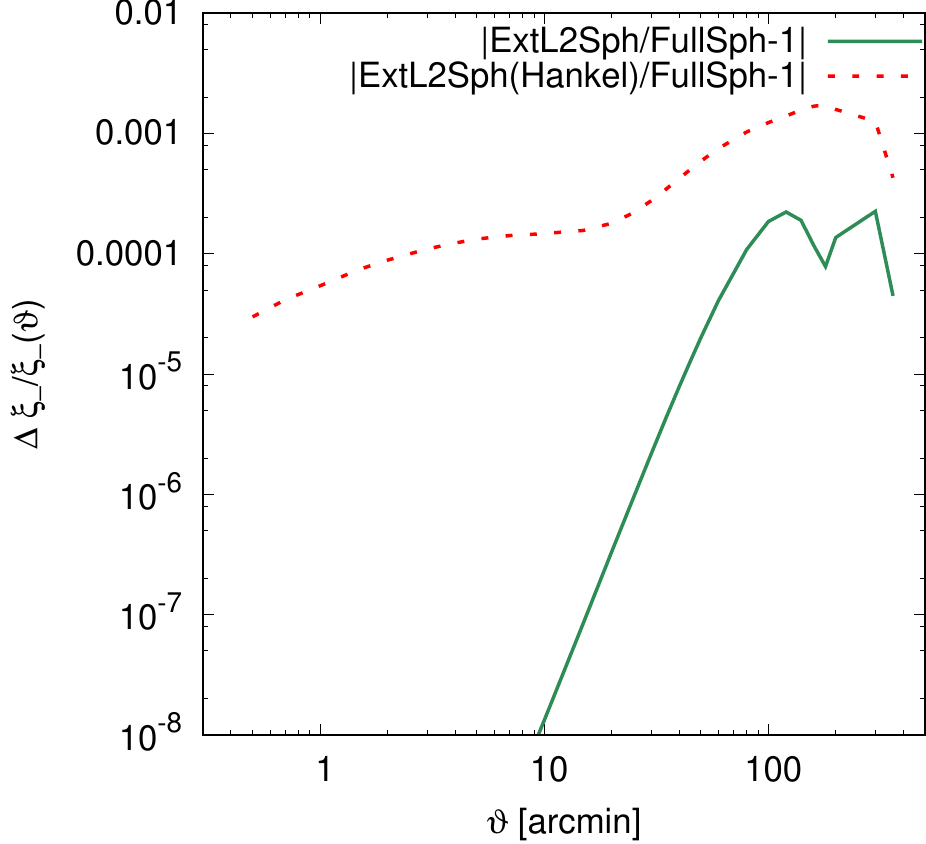}
    }
  \end{center}

  \caption{%
    The difference of the two-point shear correlation functions $\xi_+$ (\emph{left}) and $\xi_-$ (\emph{right})
    of the ExtL2Sph projection relative to the full spherical case (FullSph). Two cases of the shear correlation function transformation for
    ExtL2Spha are shown,
    the full spherical case (eq.~\ref{eq:xi_wigner}, green solid lines), and the flat-sky Hankel transform (eq.~\ref{eqn:xiGG}, red dashed curves).
  }

  \label{fig:xi_p_wigner}

\end{figure}

We compare predictions for the two-point shear correlation function using
the Hankel transformation and the full spherical transformation in
Fig.~\ref{fig:xi_p_wigner}. We show the full projection and the best
approximation, ExtL2Sph. Note that for the `FullSph' case we replace the full
projection with ExtL2Sph for $\ell > 200$ to reduce computation time.
We find that the Hankel transform (equation~\ref{eqn:xiGG}) is accurate to better than $5$ ($0.2$) percent
for $\xi_+$ ($\xi_-$).
The difference between the second-order Limber and full projection solution using the
spherical transform (equation~\ref{eq:xi_wigner}) is well below one ($0.03$) percent
on scales of $\vartheta < 300$ arcminutes for $\xi_+$ ($\xi_-$).
The red lines in Fig.~\ref{fig:xi_p_wigner} present the limit of
precision that we can achieve with the current fast Hankel transform implementation
for the correlation function. The green lines show the limit of the
second-order Limber approximation on the correlation function.

\begin{figure}

  \begin{center}
    \resizebox{1.0\hsize}{!}{
      \includegraphics{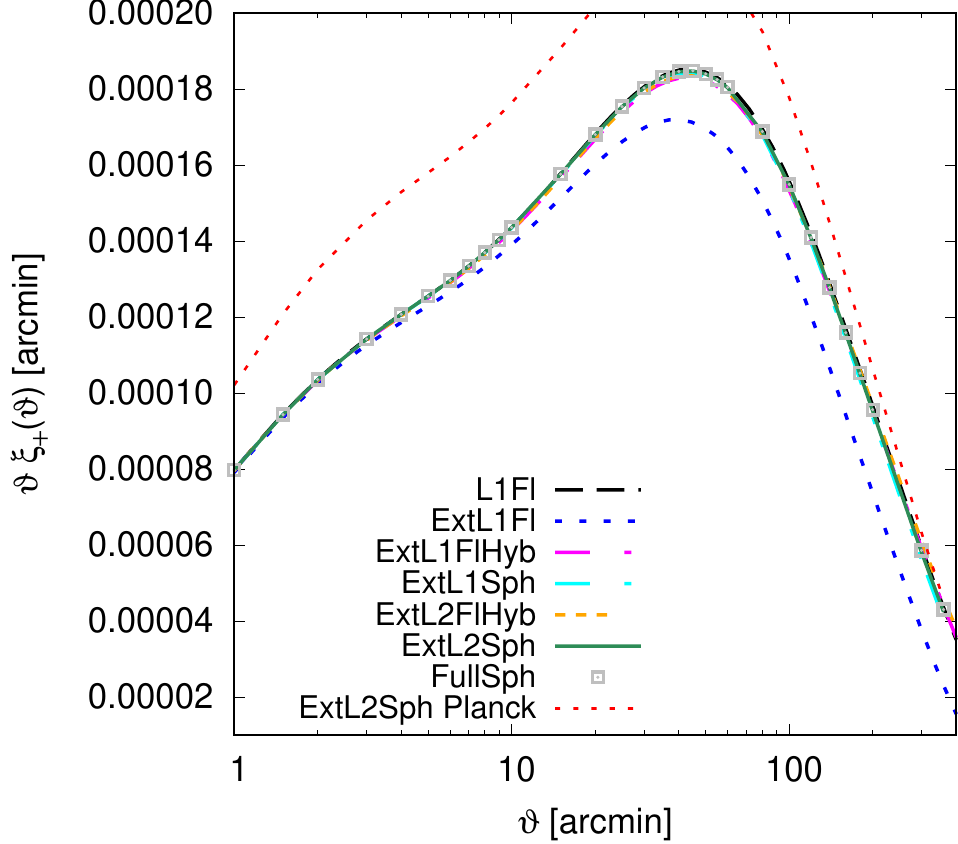}
      \hspace*{2em}
      \includegraphics{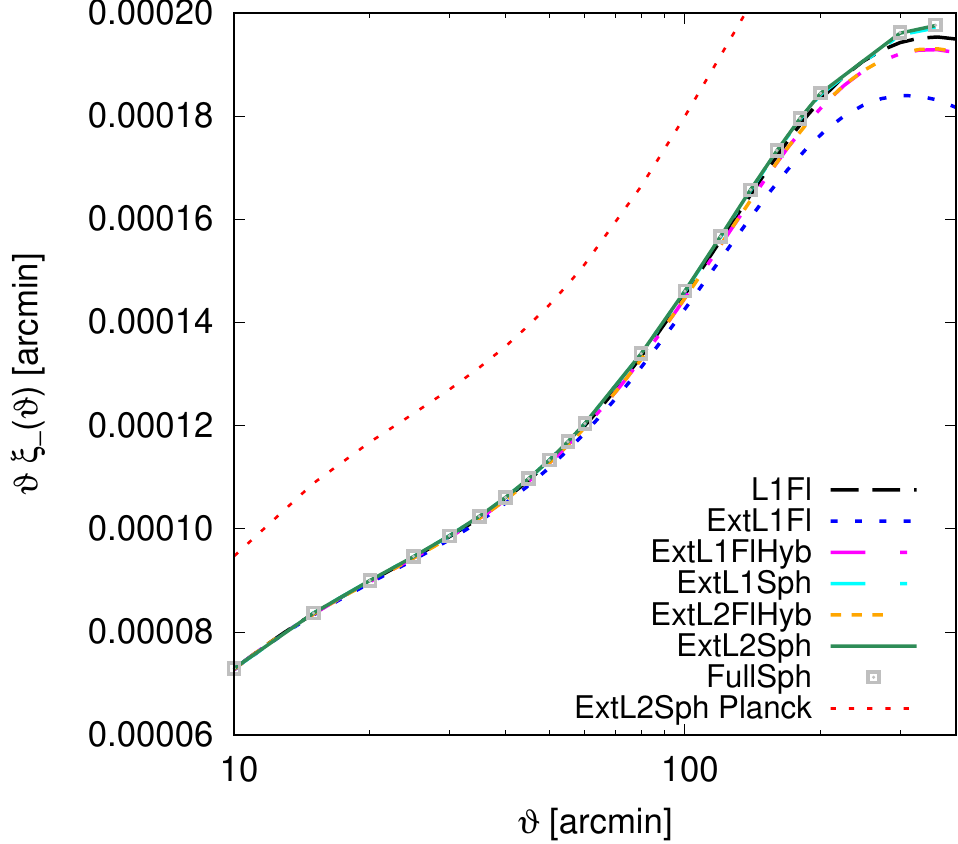}
    }

    \resizebox{1.0\hsize}{!}{
      \includegraphics{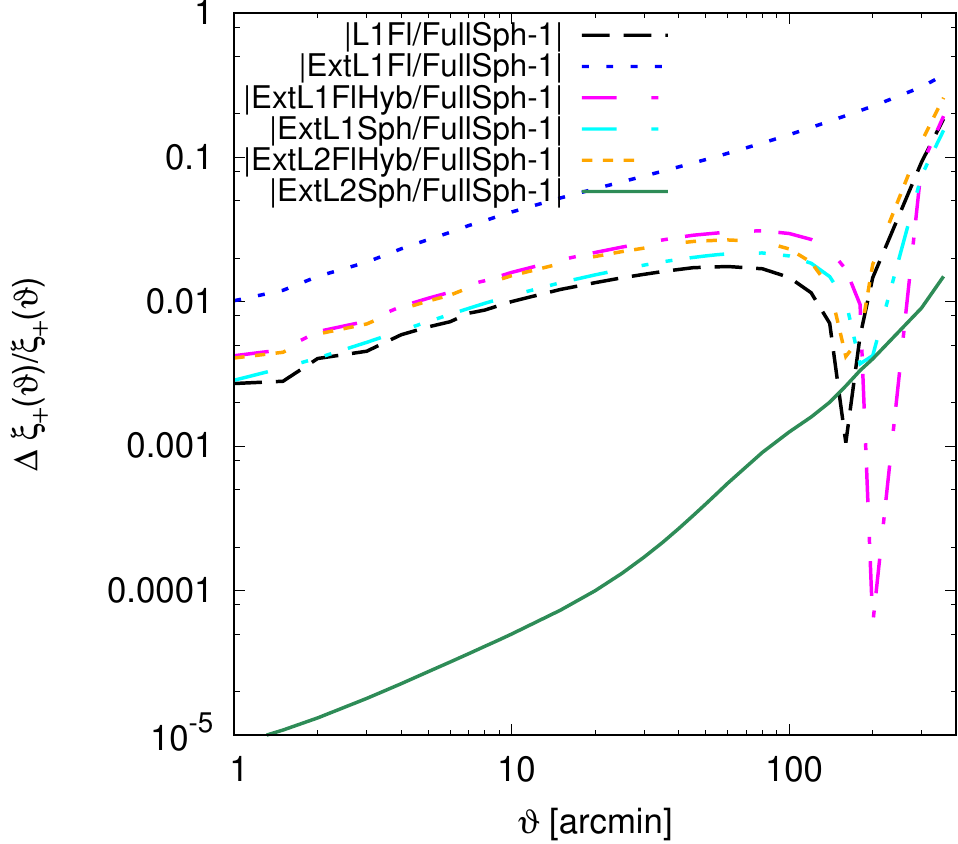}
      \hspace*{2em}
      \includegraphics{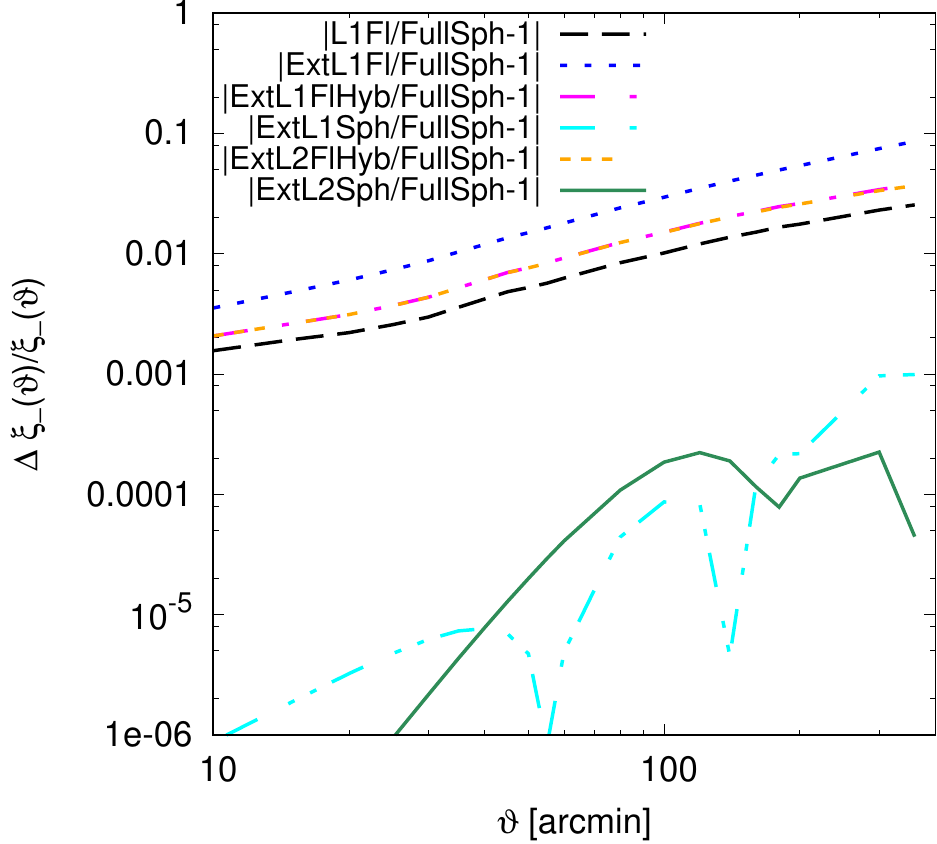}
    }
  \end{center}

  \caption{The two-point shear correlation functions $\xi_+$ (\emph{left}) and $\xi_-$ (\emph{right}).
  In the spherical cases (ExtL1Sph, ExtL2Sph, FullSph), $\xi_+$ and $\xi_-$ have been computed using the spherical
  transform (equation~\ref{eq:xi_wigner}). For the flat cases the Hankel transform
  in equation~(\ref{eqn:xiGG}) was used.
  The upper panels
  show the total shear correlation functions for the range of cases listed in Table~\ref{tab:cases}.
    The lower panels shows the relative differences to the spherical sky second-order extended Limber approximation, (ExtL2Sph).
     The theoretical models correspond to the CFHTLenS best-fit cosmological parameters with
    $\Omega_{\rm m} = 0.279$, $h=0.701$, and $\sigma_8 = 0.79$ \citep{CFHTLenS-2pt-notomo}.  For comparison we
    	also show, in the upper panels, the spherical sky second-order extended Limber approximation model for the Planck-best fit
    cosmology with $\Omega_{\rm m} = 0.3$, $h=0.67$ and $\sigma_8 = 0.83$ \citep{2015arXiv150201589P}.
  }

  \label{fig:xi_pm}

\end{figure}

Figure~\ref{fig:xi_pm} shows the two-point correlation functions $\xi_+$ (left)
and $\xi_-$ (right) using the different cases for the shear power spectrum
listed in Table~\ref{tab:cases}. The component $\xi_+$ is calculated with the
appropriate transformation, i.e.~Hankel for the flat cases, and {spherical
involving Legendre polynomials} for the spherical cases. For $\xi_-$ which is
less dominated by large scales and thus Limber and flat-sky approximations, in
comparison to $\xi_+$, we use the Hankel transform in all cases. In addition,
for $\xi_-$ the approximation `ExtL2Sph' is our reference case. The adopted
CFHTLenS redshift distribution and fiducial cosmological model are the same as
in Figure~\ref{fig:Cl_cases}. As is clear by the red dotted curve, using the
Planck cosmology \citep{2015arXiv150201589P} induces a significant change in
the amplitude of the shear correlation function in comparison to the different
projection methods.

\subsection{Application to CFHTLenS data}
\label{sec:cfhtlens}


The Canada-France-Hawaii Telescope Lensing Survey (CFHTLenS) represented a
major step forward for the field of weak gravitational lensing, in terms of
improved accuracy in data reduction \citep{CFHTLenS-data}, the implementation
of PSF-Gaussianised matched multi-band photometry
\citep{CFHTLenS-photoz}, cross-correlation clustering analysis between
photometric redshift slices to verify tomographic redshift distributions
\citep{CFHTLenS-2pt-tomo}, accurate calibrated shape measurements
\citep{CFHTLenS-shapes} and a full suite of informative systematic tests to
select a clean data set \citep{CFHTLenS-sys}. Since the public release
of this survey in 2013, the community has continued to scrutinise and advance
our understanding of CFHTLenS by identifying a number of areas where analyses
could improve:
\begin{itemize}
 \item{\citet{2016MNRAS.463.3737C} identified significant biases in the tomographic
photometric redshift distributions using a more effective clustering analysis,
in comparison to \citet{CFHTLenS-2pt-tomo}, by incorporating newly overlapping
spectroscopic data from the Sloan Digital Sky Survey.  The conclusion of this
work was that any re-analysis of CFHTLenS should include systematic error terms
to account for bias and scatter, with a prediction that accounting for these
biases would {\it reduce}\/ the recovered amplitude of $\sigma_8$ by $\sim
4$\%. Additional new techniques to calibrate the redshift distribution of tomographic
bins was introduced recently in \cite{KiDS-450}.}
\item{The CFHTLenS tomographic cosmological analysis was then revisited by
\citet{joudaki/etal:2016} in order to include a full redshift error analysis
based on the results from \citet{2016MNRAS.463.3737C}.  The impact of
correcting for these biases, including their associated errors, served to
reduce the overall constraining power of the survey and hence also the tension
between CFHTLenS and CMB constraints.}
 \item{\cite{asgari/etal:2017} used the stringent COSEBI statistic
\citep{COSEBIs} to identify significant non-lensing B-mode distortions when the
CFHTLenS data was split into tomographic slices.}
\item{\citet{2015MNRAS.454.3500K} showed that the CFHTLenS shear calibration
corrections derived in \citet{CFHTLenS-shapes} were underestimated as a result
of an imperfect match between the galaxy populations in the data and image
simulations.}
\item{\citet{2016arXiv160605337F} demonstrated that the CFHTLenS data would
have been subject to a weight bias that favours galaxies that are more
intrinsically oriented with the point-spread function.  They also showed that
the impact of calibration selection biases, that were not considered in
\citet{CFHTLenS-shapes}, would have lead to the over-correction of
multiplicative shear bias in the CFHTLenS analyses, by a few percent.}
\item{\citet{joudaki/etal:2016} updated the CFHTLenS covariance matrices using
larger-box numerical simulations that were less subject to the lack of power on
large scales. A complementary accurate estimate of the covariance matrix using
analytical methods will be published soon (Joachimi et al.~in prep.)}
\item{\cite{2012ApJ...761..152T} provided a more accurate non-linear power
spectrum correction than that used in the original CFHTLenS analyses, and the
halo model from \cite{2015MNRAS.454.1958M} allowed for simultaneous modelling
of baryonic modifications to the non-linear power spectrum.} 
\end{itemize}
All these advances in our understanding were incorporated and accounted for in
the recent KiDS cosmic shear analysis \citep{KiDS-450} which reports a $2.3
\sigma$ tension with Planck.  Efforts are now underway to fully re-analyse
CFHTLenS using the advanced KiDS analysis pipeline with revised shape
measurements and calibrations for the shear and photometric redshifts. Until
this analysis is complete we note that these known shortcomings with the
original CFHTLenS results impact in different ways the cosmological conclusions
that one can draw. As CFHTLenS has similar statistical power
to current weak lensing surveys, however, it nevertheless provides a very
useful testbed with which to demonstrate the impact of adopting different
approximations when constraining cosmological parameters.


In this work, we focus on the weak-lensing power spectrum projection, and
assess the impact of various approximations on cosmological constraints from
CFHTLenS. For consistency with the original analysis
\citep{CFHTLenS-2pt-notomo}, we adopt the same priors and non-linear power
spectrum corrections from \cite{2003MNRAS.341.1311S}.

We re-analyse the 2D CFHTLenS measurement of the two-point shear correlation
function $\xi_\pm(\theta)$ from \cite{CFHTLenS-2pt-notomo}, defined in
equation~(\ref{eqn:xiGG}). As in \cite{CFHTLenS-2pt-notomo} we fit both
components $\xi_+$ and $\xi_-$ between angular scales $\theta = 0.8$ and $350$
arc minutes, and use a $N$-body simulation estimate of the non-Gaussian
covariance including the cross-covariance between both components. Bayesian
Population Monte-Carlo parameter sampling is performed using the publicly
available software
\textsc{CosmoPMC}\footnote{\texttt{http://www.cosmostat.org/software/cosmopmc}}
\citep{WK09,KWR10}. The cosmological modelling part includes the various
lensing projections, calculated using the software library
\textsc{nicaea}\footnote{\texttt{http://www.cosmostat.org/software/nicaea}}.


For a first-order standard Limber flat-sky approximation (L1Fl) we find
$\sigma_8 (\Omega_{\rm m}/0.27)^{0.6} =0.787^{+0.031}_{-0.033}$, the same
result that was published in \cite{CFHTLenS-2pt-notomo}. Using the second-order
extended Limber flat-sky hybrid approximation (ExtL2FlHyb) results in $\sigma_8
(\Omega_{\rm m}/0.27)^{0.6} = 0.788 \pm 0.032$, a negligible change of the
amplitude that is well within the Monte-Carlo sampling noise. The largest
difference is measured with the deprecated case ExtL1Fl, for which the
recovered amplitude is larger by $16\%$ of the statistical error. These
negligible changes to the error bars were to be expected owing to the high
level of statistical noise and cosmic variance in comparison to the low-level
impact of the various approximations shown in Fig.~\ref{fig:Cl_cases}.

Table~\ref{tab:CFHTLenS_Sigma8} lists the mean and 68\% credible interval for
$\sigma_8 \Omega_{\rm m}^{0.6}$ for the various approximations to the
lensing power spectrum projections listed in Table~\ref{tab:cases}. Note again
that these values do not represent the state-of-the-art cosmological results,
since many of the above listed analysis advancements made since 2013 have not
been taken into account. As an example of a significant effect, when using the
revised non-linear power spectrum of \cite{2012ApJ...761..152T} in place of
\cite{2003MNRAS.341.1311S}, there is a decrease of $0.6 \sigma$ with $\sigma_8
(\Omega_{\rm m}/0.27)^{0.6} = 0.768^{+0.029}_{-0.031}$, using L1Fl.

Considering the cosmological constraints from tomographic Kilo-Degree Survey
(KiDS), we conclude that these are robust to flat-sky and Limber
approximations. The case ExtL1FlHyb that was used for the analysis of KiDS data
in \citet{KiDS-450} and \cite{joudaki/etal:2017} introduces errors that are
more than an order of magnitude lower than the cosmic variance for that survey,
and thus this approximation has a negligible impact on the cosmological
parameters.

\renewcommand{\baselinestretch}{1.5}
\begin{table}
\begin{centering}
  
  \caption{\label{tab:CFHTLenS_Sigma8}Mean and 68\% credible interval for 
  $\sigma_8 (\Omega_{\rm m}/0.27)^{0.6}$ and $\sigma_8 (\Omega_{\rm m}/0.3)^{0.6}$
  for various approximations to the lensing
  power spectrum projections listed in Table~\ref{tab:cases}.}

  \begin{tabular}{lcc} \hline
  ID         & $\sigma_8 (\Omega_{\rm m}/0.27)^{0.6}$ & $\sigma_8 (\Omega_{\rm m}/0.3)^{0.6}$ \\ \hline
  L1Fl       & $0.787^{+0.031}_{-0.033}$ & $0.739^{+0.029}_{-0.031}$ \\
  ExtL1Fl    & $0.792 \pm 0.032$ & $0.744 \pm 0.030$ \\
  ExtL1FlHyb & $0.788^{+0.031}_{-0.033}$ & $0.740^{+0.029}_{-0.031}$ \\
  ExtL2FlHyb & $0.788^{+0.031}_{-0.033}$ & $0.740^{+0.029}_{-0.031}$ \\
  ExtL2Sph(Hankel) & $0.789^{+0.031}_{-0.032}$ & $0.740^{+0.029}_{-0.030}$ \\ \hline
  \end{tabular}

\end{centering}
\end{table}
\renewcommand{\baselinestretch}{1}

\subsection{Alternative two-point shear statistics; the mass aperture statistic
and COSEBIs}

\begin{figure}
\begin{center}
\begin{tabular}{ccc}
\includegraphics[width=0.60\textwidth]{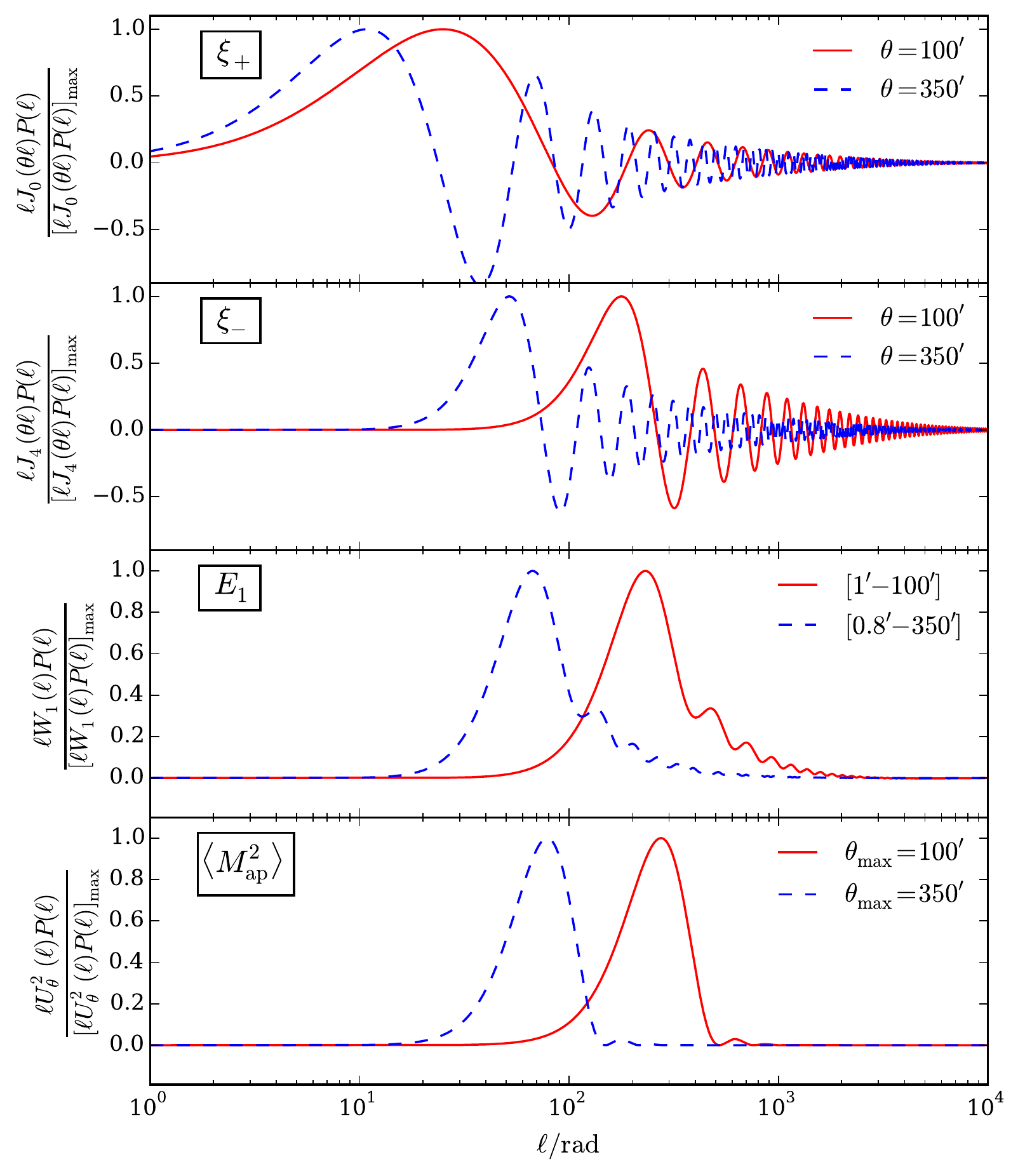} \\
\end{tabular}
\caption{\small{\label{fig:filters} 
Integrand of $\xi_+$ (upper), $\xi_-$
(upper middle), $E_1$ (lower middle, E-COSEBIs) and $\langle M_{\rm ap}
\rangle^2$ (lower panel). All integrands are of the form $\ell F(\ell)
P(\ell)$, where $F(\ell)$ is the corresponding weight-function for each
statistic and $P(\ell)$ is the E-mode convergence power spectrum, with the
exception of $\xi_\pm$, for which $P(\ell)$ is equal to the sum of the E and
B-mode power spectra. Two cases are shown for each statistic as listed in each
caption. For the aperture mass statistic $\theta_{\rm max}=2\theta$ is shown.
Note that higher order COSEBIs modes generally probe larger $\ell$-modes, hence
here we only show the lowest mode $E_1$. All values are normalized with respect
to their maximum value. This figure illustrates how different two-point cosmic shear
statistics have different dependences between the angular scales sampled and the $\ell$-range probed. }}
\end{center}
\end{figure}

The two-point shear correlation function $\xi_\pm$ represents the current
baseline observable for cosmic shear measurements.   As shown in
Figure~\ref{fig:xi_pm}, however, using the standard first-order extended
Limber flat-sky approximation (ExtL1FlHyb) can result in errors exceeding 10
percent, on angular scales $\theta > 300$ arcmin.    This is a result of the
weight given to low $\ell$ modes in the $\xi_+$ statistic, as
illustrated in Figure~\ref{fig:filters} which shows the integrand of $\xi_+$
and $\xi_-$ (upper two panels) for two cases ( $\theta = 100$ and $\theta =
350$ arcmin), normalised to their maximum value.   This error does not impact
CFHTLenS analyses, given the low signal-to-noise of the measurements on these
scales.  It will however become increasingly important for upcoming wider-field
surveys that will accurately probe these scales.

In this paper we provide a solution in the form of the second-order extended
Limber approximation, but another option to consider is the use of alternative two-point
shear statistics that are less sensitive to accuracy in shear power spectrum
measurement at low $\ell$.   Both the aperture-mass dispersion, $\langle M_
{\rm ap}^2 \rangle$ \citep{1998MNRAS.296..873S}, and the Complete Orthogonal
Sets of E/B-mode Integrals (COSEBIs), $E_n$ \citep{COSEBIs} statistics satisfy
this requirement and are linearly related to the shear power spectrum in the
flat-sky approximation via integrals of the form 
\begin{align}
  \langle M_ {\rm ap}^2 \rangle(\theta) & = \frac 1 {2\pi} \int_0^{\infty}\d \ell \, \ell \,
  \hat U^2(\theta\ell) P^\gamma(\ell),
  \label{eqn:integ}
  \\
  E_n & = \frac 1 {2\pi} \int_0^{\infty}\d \ell \, \ell \, W_n(\ell) P^\gamma(\ell),
\end{align}
where the Fourier-space filter functions $\hat U$ and $W_n$ are defined in
\cite{1998MNRAS.296..873S} and \cite{COSEBIs}, respectively.
Figure~\ref{fig:filters} shows the integrands of these statistics, again
normalised to their maximum value, where the integrands are of the form $\ell
F(\ell) P^\gamma(\ell)$.  The lower middle panel in Figure~\ref{fig:filters}
shows the COSEBIs integrands for two angular ranges, $[1',100']$ and
$[0.8',350']$, where we only show the integrands for the lowest COSEBIs mode,
$E_1$, as the higher modes generally probe larger $\ell$-modes.  The lowest
panel shows the integrands of aperture mass dispersion statistics, for the same
two maximum angular ranges. 

Note that the development of the aperture-mass dispersion statistic, $\langle
M_ {\rm ap}^2 \rangle$ was initially motivated to enable the separation of the
measured signal into an E-mode (cosmological signal) and B-mode (systematics).
This statistic is, however, a lossy conversion and is biased by small angular
separations, where blending of galaxies makes shear measurement challenging
\citep{KSE06}. The COSEBIs statistic tackles both these shortcomings.
\citet{CFHTLenS-2pt-notomo} present a detailed comparison of cosmological
constraints obtained from this range of different two-point shear statistics
finding consistent results.

\section{Conclusions}
In this paper we evaluate precision theoretical calculations for cosmic shear
observables, bringing together different sources from the literature to provide
a pedagogical review of the impact of adopting flat-sky and Limber
approximations.  We demonstrate that for current surveys, such as CFHTLenS
and KiDS, these approximations have a negligible impact on cosmological
parameter constraints.

For future surveys, the decrease in statistical errors places higher
requirements on the accuracy of the theoretical modelling.    There is also,
however, the need to be able to rapidly sample the multi-dimensional
cosmological parameter likelihoods.  This requirement for computational speed
is incompatible with a theoretical analysis that calculates a full spherical
solution for the shear power spectrum, without adopting any approximation.  We
therefore present alternative solutions, revisiting the work of
\citet{2012PhRvD..86b3001B} who showed that adopting the second-order extended
Limber approximation of \citet{2008PhRvD..78l3506L} provides a representation
of the full spherical solution for the shear power spectrum that is accurate at
the sub-percent level for $\ell > 3$.    We have verified this result and
provide to the community our fast numerical implementation of all the
approximations studied in this analysis, and the slow calculation of the full
projection within the publicly available package \textsc{nicaea} at
\texttt{http://www.cosmostat.org/software/nicaea}.

Finally we propose that future surveys seek to optimise the statistical
analyses of their cosmic shear data.  For example moving from the standard
two-point shear correlation function statistic to the more stringent `COSEBI'
statistic \citep{COSEBIs} renders the cosmic shear measurement insensitive to
the low-$\ell$ scales where the Limber and flat-sky approximations have an
impact on the precision of the theoretical modelling.  

We have considered a flat Universe throughout this paper. To generalise
the calculations to non-flat models, one needs to modify the comoving angular
diameter distance to account for the spatial curvature $K \ne 0$. In addition,
the spherical Bessel functions are replaced with hyperspherical Bessel
functions \citep{1986ApJ...308..546A}. In a universe with positive curvature
$K>0$ the 3D wave modes become discrete integer variables. For non-flat models,
we do however not expect qualitative differences from our results.

\section*{Acknowledgments}

The authors thank Tommaso Giannantonio, Sarah Bridle, Ami Choi, Donnacha Kirk,
Lance Miller, Benjamin Joachimi, Chris Blake, Joe Zuntz, Joanne Cohn, Alex Hall and Adam Amara for very helpful
discussions. This work was financially supported by the DFG (Emmy Noether grant
Hi 1495/2-1; SI 1769/1-1; TR33 `The Dark Universe'), the Alexander von Humboldt
Foundation, the ERC (grants 279396, 647112), the Seventh Framework Programme of
the European Commission (Marie Sklodwoska Curie Fellowship grant 656869), the
Netherlands Organisation for Scientific Research (NWO) (grants 614.001.103),
NSERC, CIfAR, and the World Premier International Research Center Initiative
(WPI), MEXT, Japan. Parts of this research were conducted by the Australian
Research Council Centre of Excellence for All-sky Astrophysics (CAASTRO),
through project number CE110001020i, and by the German Federal
Ministry for Economic Affairs and Energy (BMWi) under
project no. 50QE1103.

\bibliographystyle{mnras}
\bibliography{astro}

\begin{appendix}

\section{Derivation of the weak-lensing power spectra}
\label{sec:derivations_C}

The following derivations are detailed in \cite{2000PhRvD..62d3007H} and
\cite{2005PhRvD..72b3516C}, and are provided here for completeness.

\subsection{Spherical case}

\subsubsection{Lensing potential power spectrum}

To obtain the power spectrum of the lensing potential, 
we insert the lensing projection (equation~\ref{eq:psi}) into the
inverse harmonics expansion (equation~\ref{eq:psi_harm_exp_inv}) and write the 3D potential
as its Fourier transform (equation~\ref{eq:hatPhi_inv}) to get
\begin{equation}
  \psi_{\ell m} = \frac 2 {c^2} \int {\rm d} \Omega {\rm Y}^\ast_{\ell m}(\theta, \varphi)
    \int_0^\infty \frac{{\rm d}\chi}{\chi} q(\chi) \int \frac{{\rm d}^3 k}{(2\pi)^3} \hat \Phi(\vec k; \chi) {\rm e}^{-{\rm i} \vec k \cdot \vec r}.
\end{equation}
The 3D position vector $\vec r$ is a 3D position vector with polar coordinate
$r = \chi$ and polar angles $(\theta, \varphi)$. Similarly we denote with
$\theta_k, \varphi_k$ the polar angles of the 3D Fourier vector $\vec k$. We
insert the expansion of a plane wave into spherical harmonics,
%
%
\begin{equation}
  {\rm e}^{{\rm i} \vec k \cdot \vec r} = 4 \pi \sum_{\ell=0}^{\infty} \sum_{m=-\ell}^{\ell}
    {\rm i}^\ell \, {\rm j}_\ell(k \chi)
    {\rm Y}_{\ell m}(\theta, \varphi) {\rm Y}_{\ell m}(\theta_k, \varphi_k) .
  \label{eq:wave_exp}
\end{equation}
Making use of the orthogonality of the spherical harmonics
%
%
\begin{equation}
  \int {\rm d} \Omega {\rm Y}^{}_{\ell m}(\theta, \varphi) {\rm Y}^\ast_{\ell^\prime m^\prime}(\theta, \varphi) = \delta_{\ell \ell^\prime} \delta_{m m^\prime},
  \label{eq:Yellm_ortho}
\end{equation}
the expression for $\psi_{\ell m}$ simplifies to
\begin{equation}
  \psi_{\ell m} = \frac {{\rm i}^\ell}{c^2 \pi^2} \int_0^\infty \frac{{\rm d}\chi}{\chi} q(\chi) \int {\rm d}^3 k \,
    \hat \Phi(\vec k; \chi) {\rm j}_\ell(k \chi) {\rm Y}_{\ell m}(\theta_k, \varphi_k).
  \label{eq:psi_ellm}
\end{equation}
To obtain the potential power spectrum, we take the absolute square of the
last equation and use the definition of the 3D potential power spectrum
(equation~\ref{eq:p_phi}). The delta-function resolves one 3D Fourier integral.
We split the second integration into radial and spherical coordinates, ${\rm
d}^3 k = {\rm d} k k^2 {\rm d} \Omega_k$ and use once again the orthogonality
of the spherical harmonics to resolve the spherical integral. This leads to the
potential power spectrum in equation~(\ref{eq:C_ell_phi_Pphi}).

\section{Discussion and comparison to previously published work}
\label{app:B}

In this section we briefly discuss previously published work on the full
projection and second-order Limber equation of the lensing power spectrum, cross-checking and comparing their
results with our independent findings.

\subsection{Kitching et al.~2016 (version 1)}

\cite{2016arXiv161104954K} compute the full projection of the weak-lensing
power spectrum, which they present as spherical-radial representation of the 3D
shear field. Our results in equation~(\ref{eq:C_ell_full}) corresponds to their
Equations (7) and (8) assuming a flat Universe and the case of perfect
photometric redshifts, with $p(z | z_{\rm p}) = \delta_{\rm D}(z - z_{\rm p})$,
and for a bin function that selects the redshift bin of $z$, $W^{\rm SR}(z,
z_{\rm p})$ is unity if $z_{\rm p}$ is in the redshift bin denoted by $z$,
and zero otherwise. We find that Equation (7) in \cite{2016arXiv161104954K} is
missing a factor $2/\pi$.

\cite{2016arXiv161104954K} derive the spherical and extended Limber
approximation starting from the full spherical projection in their Appendix.~A.
We find that the filter function $q$ defined in their Equation (31) has an
additional factor of comoving distance $r$, and an additional factor of
$\pi/2$.

As shown in this paper, we are unable to reproduce the differences that
\cite{2016arXiv161104954K} report, between the full spherical solution and the
different approximations, neither for the power spectrum nor for the shear
correlation function.

\subsection{Bernardeau, Bonvin, Van de Rijt, Vernizzi (2012); Van de Rijt (2012)}

\cite{2012PhRvD..86b3001B} present the non-tomographic full projection
$C(\ell)$ in the approximation of the 3D potential power spectrum $P_\Phi$
separating into $k$- and $\chi$-dependent functions in their Equation (44).
Their expression holds for a single source redshift.

The PhD thesis of \cite{vande2012} presents an explicit calculation of the
second-order Limber approximation. They carry out the derivatives of the
kernels $f$ under the assumption of a constant growth-suppression factor
$D_+(a)/a$. Then, for a constant source comoving distance $\chi_{\rm S}$, the
lensing efficiency in equation~(\ref{eq:lens_efficiency}) is $q(\chi) =
(\chi_{\rm S} - \chi) \chi_{\rm S}^{-1}$, and the derivative of the separated
kernel function in equation~(\ref{eq:f_LA08_s}) can be calculated analytically,
\begin{equation}
  f_{\rm s}^{\prime\prime}(\chi) + \frac{\chi}{3} f_{\rm s}^{\prime\prime\prime}(\chi)
    = - \frac 1 8 \chi^{-5/2} \left( \frac 5 \chi
          + \frac 1 {\chi_{\rm S}} \right) \frac{D_+(\chi)}{a(\chi)}.
\label{eqn:B1}
\end{equation}
This result confirms the power-law behaviour for $\chi \ll \chi_{\rm S}$ of the
filter function, which we exploited earlier to fit this function.

Inserting equation~(\ref{eqn:B1}) into the second-order Limber power spectrum
in equation~(\ref{eq:C_ell_limber2_dr_s}) and using the inverse Poisson
equation to replace the matter with the potential power spectrum we obtain the
same expression as \cite{vande2012} (their Equation~7.19).

\begin{figure}

  \begin{center}
    \resizebox{0.5\hsize}{!}{
      \includegraphics{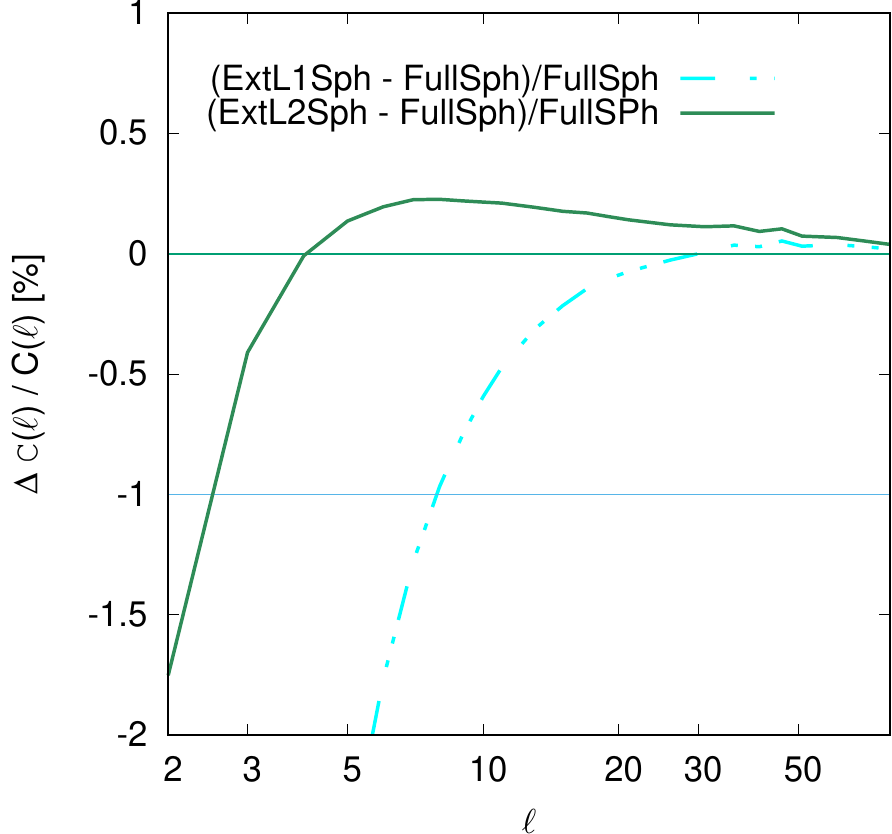}
    }
  \end{center}

    \caption{The relative differences in percentage of the spherical first- and second-order Limber
    shear power spectra with respect to the full projection
    as function of wave mode $\ell$, (see Table~\ref{tab:cases}). In this figure, the redshift distribution is chosen to be
    at a single source plane at $z_{\rm S}=1$, with the cosmological parameters (see text) are chosen to match
    \citet{vande2012}, see their Figure~7.3 for comparison.
    }

    \label{fig:L1L2E_Rijt}

\end{figure}

In Fig.~\ref{fig:L1L2E_Rijt} we reproduce Fig.~7.3 from \cite{vande2012} using
a similar set up, a flat $\Lambda$ Universe with $\Omega_{\rm m} = 0.3, h=0.65,
\Omega_{\rm b} = 0.0461, \sigma_8 = 0.8, n_{\rm s} = 0.96$. All source galaxies
are at redshift $z_{\rm S} = 1$. The non-linear 3D matter power spectrum from
\cite{2012ApJ...761..152T} is used. The ratio of the first-and second-order
Limber approximated power spectra to the full projection shows excellent
agreement at the sub-percent level.

\subsection{LoVerde \& Afshordi (2009)}

This paper introduces the extended Limber approximation to second order that we
apply in this work. Although they present the specific case of 2D galaxy
clustering, their calculations are general enough to apply to a weak-lensing
context. Their Equation (5) is a spherical cross-power spectrum of two scalar
fields $A$ and $B$, projected from 3D to 2D via projection kernels $F_A$ and
$F_B$ defined in their Equation (4). Comparing their expressions with the
weak-lensing potential (\ref{eq:psi}), we set $F_A(\chi) = F_B(\chi) = 2 c^{-2}
D_+(\chi) q(\chi) \chi^{-1}$. With that, their Equation~(5) is identical to
equation~(\ref{eq:C_ell_phi_Pphi}).

The second-order Limber approximation in \cite{2008PhRvD..78l3506L} is
presented in eq.~(12). This is consistent with our first-order
(equation~\ref{eq:C_ell_limber1}) and second-order
(equation~\ref{eq:C_ell_limber2_dr}) Limber approximation terms, when
accounting for the difference between lensing potential and shear 2D power
spectrum, and 3D potential and matter power spectrum.

\subsection{Schmidt (2009)}
\label{sec:schmidt08}

\cite{2008PhRvD..78d3002S} derive the lensing power spectrum in the flat-sky
limit, see their Equation (9). Inserting the Poisson equation and the
growth function, and writing the redshift filter function
$W_\kappa$ (eq.~10) in terms of the lensing efficiency and comoving distances,
$W_\kappa[z(\chi)] = H(z)^{-1} \chi q(\chi)$, their expression, using our
notations, reads
\begin{align}
  P_{ij}^\gamma(\ell) = & \frac 2 \pi \, \pref^2
                 \int_{0}^\infty {\rm d} \chi \chi \, \frac{q_i(\chi)}{a(\chi)}
                \int_{0}^\infty {\rm d} \chi^\prime\, \chi^\prime
                \frac{q_j(\chi^\prime)}{a(\chi^\prime)}
                \int_0^\infty {\rm d} k \, k^2 \, P_{\rm m}(k; \chi, \chi^\prime) \,
                {\rm j}_\ell(k \chi) \, {\rm j}_\ell(k \chi^\prime) ,
  \label{eq:P_ell_gamma}
\end{align}
This is consistent with our equation (\ref{eq:P_ell_gamma_full_der}) under the
additional assumption that mainly modes with $k \chi \approx k \chi^\prime
\approx \ell$ contribute to the integral; that is modes around the maxima of
the Bessel functions. Then, we can draw out of the integral the factor $\ell^4
\approx k^4 \chi^2 {\chi^\prime}^2$, to recover
equation~(\ref{eq:P_ell_gamma_full_der}).
In fact, using the approximation $k \chi = \ell$ seems to go too far, this is
already halfway the Limber approximation.

\subsection{Giannantonio et al. (2012)}
\label{sec:giannantonio12}

\cite{2012MNRAS.422.2854G} derive the flat-sky lensing power spectrum in their
Equations (25) and (26). Their window function $W^{\varepsilon_i}$ defined in
their Equation (25) for a flat Universe equals $\pref q(\chi) / a(\chi)$, since
${\rm d} z \, ({\rm d} N(z)/{\rm d} z) = {\rm d} \chi n(\chi)$; however due to
a typo there is a factor $r_K[r(z)]$ missing in the window function, which
translates into a missing $r_K[r] r_K[r^\prime]$ in the full projection
integral\footnote{Typo confirmed by T.~G., priv.~comm.}. This also leads to an
errornous $r_K^{-2}(r)$ in their Limber equation (27). With these factors
accounted for, and making the additional approximation $k \chi \approx k
\chi^\prime \approx \ell$ (see their Appendix~\ref{sec:schmidt08}) we reproduce
the expressions of \cite{2012MNRAS.422.2854G}.

\section{Fast evaluation of the shear correlation function on the sphere}
\label{app:C}

The calculation of the shear correlation function on the sphere requires the
estimation of the reduced Wigner ${\rm D}$-matrices ${\rm d}^\ell_{2\,2}$ and
${\rm d}^\ell_{2\,-2}$. The general calculation of ${\rm d}^\ell_{m\,n}$ is
cumbersome, but there are quick and numerically stable recurrence relations if
we are interested in only a subset of these matrices. In particular, following
\citet{BLANCO199719} we can show that
\begin{equation}
{\rm d}^\ell_{m\,n}={\ell(2\ell-1)\over \sqrt{[\ell^2-m^2][\ell^2-n^2]}}
\left[ \left( {\rm d}^1_{0\,0}-{m\,n\over \ell(\ell-1)} \right) {\rm d}^{\ell-1}_{m\,n}
-{\sqrt{[(\ell-1)^2-m^2][(\ell-1)^2-n^2]}\over (\ell-1)(2\ell-1)} {\rm d}^{\ell-2}_{m\,n}\right] ,
\end{equation}
which allows to calculate all required reduced Wigner ${\rm D}$-matrices from the first two elements.

\label{lastpage}

\end{appendix}

\end{document}